\documentclass[journal]{IEEEtran}
\usepackage{cite}
\usepackage{lipsum}
\usepackage{multicol}
\ifCLASSINFOpdf
 
\else
 
\fi

\usepackage{subfig}
\usepackage{color}

\usepackage{hyperref}

\usepackage{amsmath,epsfig,graphics,amssymb}
\usepackage{amsmath}
\usepackage{amsfonts}
\usepackage{amssymb}
\usepackage{amsfonts,amsmath,amssymb,epsfig}
\graphicspath{{figures/}}
\begin{document}

\title{On M-ary Distributed Detection for Power Constraint Wireless Sensor Networks }
%
%
%

\author{Zahra~Hajibabaei,
        Jalil~Modares,~\IEEEmembership{Student Member,~IEEE},
        Azadeh~Vosoughi,~\IEEEmembership{Senior Member,~IEEE},
\thanks{Z. Hajibabaei and J. Modares are with the EE Department, University at Buffalo, Buffalo,
NY, e-mail: zahrahaj@buffalo.edu, jmod@buffalo.edu.}
\thanks{A. Vosoughi is with the EECS Department, University of Central Florida, Orlando, FL, e-mail:azadeh@ucf.edu.}}


\maketitle

\begin{abstract}
We consider a wireless sensor network (WSN),
consisting of several sensors and a fusion center (FC), which is
tasked with solving an M-ary hypothesis testing problem. Sensors
make M-ary decisions and transmit their digitally modulated
decisions over orthogonal channels, which are subject to Rayleigh
fading and noise, to the FC. Adopting Bayesian optimality
criterion, we consider training and non-training based distributed
detection systems and investigate the effect of imperfect channel
state information (CSI) on the optimal maximum a posteriori
probability (MAP) fusion rules and optimal power allocation between sensors, when
the sum of training and data symbol transmit powers is fixed. We consider J-divergence criteria to do power allocation between sensors. The theoretical results show that J-divergence for coherent reception will be maximized if total training power be half of total power, however for non coherent reception, optimal training power which maximize J-divergence is zero. The simulated results also show that probability of error will be minimized if training power be half of total power for coherent reception and zero for non coherent reception.  

\end{abstract}


\IEEEpeerreviewmaketitle

\section{Introduction}
\label{sec:intro}
We consider a wireless sensor network (WSN), consisting a set of spatially distributed sensors and a fusion center (FC), that is tasked with solving an $M$-ary distributed detection problem. In particular, we consider the problem of distributed classification of $M$ independent Gaussian sources with identical variances and different means. Sensors make local decisions individually and transmit their digitally modulated decisions to the FC, over orthogonal fading channels. The FC is tasked with fusing all the received signals from the sensors directly, via applying the optimal fusion rule, and making the final decision.\\
Channel-aware binary distributed detection for fusion of binary decisions transmitted over fading channels was first discussed in \cite{biaochen2006}, where the FC fuses the received signals from the sensors directly (without demodulating the transmitted symbols). They considered different schemes to improve detection performance in distributed detction systems. In \cite{Jiang}, massive MIMO has been investigated in WSN's for coherent channel.
In \cite{Varsh}, They consider a censoring sensor network scheme and show that by adding artificial noise, the system performs very closely to the exact copula-based GLRT. In \cite{Braca}, they showed that using only one transmission, the detection error can be made as small as desired. In \cite{Nad}, they considered the problem of designing binary sensor quantizers that maximize the Kullback-Leibler (KL) divergence at the fusion center (FC)
The works on channel-aware binary distributed detection are mainly built on the assumption that perfect knowledge of phase or amplitude of the fading channel coefficients are available at the FC \cite{vincentpoor,multipleantennamac}. 
Today's wireless communication systems with coherent reception rely upon training in order to facilitate channel estimation at the receiver. In fact, quantifying the effect of imperfect channel state information (CSI) and channel estimation error on the design and performance of wireless communication systems is a challenging problem, that has attracted the attention of researchers over the past decade \cite{channelestimation}. Recently, channel-aware binary distributed detection with imperfect CSI was studied in \cite{6516873,confhamid,nevat}.In \cite{nevat}, when sensors amplify and forward their observation, the performance of system under different knowledge of channel has been investigated. Sensors can send their information to the FC over Multiaccess channels or parallel access channels. In \cite{pacvsmac}, the performance of system MAC Vs PAC for Non coherent reception has been studied. in \cite{Li}, the performance of system when channel is non coherent has been investigated and it has been shown on-off keying scheme is the most energy efficient modulation scheme when the channel is subject to Rayleigh fading. In \cite{Ciu}, the optimality of the received energy test in scenarios with correlated sensor decisions and non identical sensors has been verified. In \cite{Maleki}, they proposed cooperative fusion architecture which enhance detection performance.
All of these works are for the case that we have just two hypotheses and the FC should make decision between two hypotheses. channel-aware $M$-ary distributed detection when the communication channels are modeled as additive white Gaussian noise (AWGN) \cite{1413463} and Rayleigh fading with perfect CSI available at the FC has been studied in  \cite{1413463, nahal,blindalgorithm}. In \cite{nahal} the impact of imperfect CSI on the design and performance of channel-aware $M$-ary distributed detection systems has been investigated.\\
Due to limitation of power in WSN's, it has been attempted to improve the performance of system by optimal power allocation between sensors. In \cite{Nur}, Sensors share the quantized data of their observation and it has been shown power consumption is 50 percent less than unquantized conventional methods. In \cite{Maya}, sensors transmit different linear combinations of their measurements through a multiple access channel (MAC) to FC and show their scheme save energy in the low signal-to-noise ratio regime. In \cite{Naj}, Sensors are selected based on satisfying the average global probability of detection and minimizing the energy consumption. In \cite{vincentpoor}, the power allocation between sensors for binary hypotheses and known channel fading coefficients based on cost function J divergence has been done. In \cite{deflection}, power allocation based on maximization of deflection coefficient for partial coherent and channel statistic has been done. In \cite{goodman}, local power control strategy for multi-hypotheses has been introduced.The works on power allocation in distributed detection are mainly built on the assumption that perfect knowledge of the fading channel coefficients are available at the FC and power allocation for channel estimation has not been considered.In \cite{cihan} power allocation for data and training of distributed estimation has been studied. In \cite{cihan} power allocation of data and training of distributed estimation of a source when sensors amplify and forward their observation over MAC. In \cite{Chaudhary}, power allocation between sensors when sensors amplify and forward their spatially corrolated date to FC when we have perfect channel or estimation of channel gain. To the best of our knowledge, there is no work in distributed detection for power allocation of data and training between sensors.
In order to do power allocation between sensors, we need a cost function. Deriving probability of error at the FC is hard even for the case when we know complete information about the amplitude and phase of channel. We choose J divergence as our optimality criterion. \cite{goodman} established a lower bound on pairwise sum of the individual error probabilities between two hypotheses, where the lower bound is a constant (determined by the a priori probabilities of the hypotheses) minus $J_{tot}$. Hence, by maximizing $J_{tot}$ we minimize the lower bound on the error probability."

%
%
In this work, we address the following questions: how should the power allocated between sensors to improve performance? how can we mitigate the negative impact of channel estimation error via optimizing transmit power allocation between data and training symbols? how do the answers to the above questions change as the reception mode at the FC and modulation scheme at the sensors vary? For non coherent reception, how does the power allocation differ for training and non-training based systems, where the sensors do not transmit training symbols (for estimating channel amplitudes) and the FC only relies on the knowledge of the channel statistics?
To answer these questions we consider the following three cases, assuming Rayleigh block fading channel model: ($i$) the FC is equipped with a coherent receiver and a training based channel estimator, sensors employ $M$-PSK modulation for transmitting their data and training symbols, ($ii$) the FC is equipped with a non coherent receiver and a training based channel amplitude estimator, the sensors employ $M$-FSK modulation for transmitting their data and training symbols, ($iii$) the FC is equipped with a non coherent receiver without a channel estimator (the FC only has the channel statistics), the sensors employ $M$-FSK modulation for transmitting their data symbols. The organization of the paper follows: Section II introduces the system model. Sections III and IV derive the optimal power allocation for cases ($i$),($ii$),($iii$) explained above. Section V includes our numerical results. Section VI concludes the paper.\newline

\section{System Model and Problem Statement}
\label{sec:format}
We consider the problem of testing which of the $M \geq 2$ hypotheses
 $\left\lbrace \mathcal{H}_m\right\rbrace_{m=1}^{M} $ has been occurred, assuming $\pi_m$ is the {\it a priori} probability that hypothesis $\mathcal{H}_m$ occurs. Our system model consists of a FC and $N$ spatially distributed sensors, which is tasked with solving this $M$-ary hypothesis testing problem. Let $\boldsymbol{x}_k$ denote the local observation at sensor $k$ during an observation period. We assume that $\boldsymbol{x}_k$'s are independent across sensors, conditioned on a particular hypothesis,  i.e.,  $f(\boldsymbol{x}_1,\boldsymbol{x}_2,...,\boldsymbol{x}_N|\mathcal{H}_m)=\prod_{k=1}^{N}f(\boldsymbol{x}_k|\mathcal{H}_m)$ for $m=1,...,M$, where $f(.)$ denotes the probability density function (pdf).  Suppose $\boldsymbol{x}_k$ at sensor $k$  under hypothesis $\mathcal{H}_m$ is:
\begin{equation}\label{sensing-model}
\mathcal{H}_m: \boldsymbol{x}_k=\boldsymbol{z}_m^k+\boldsymbol{\nu}_{k},  ~~~~ m=1,...,M, k=1,...,N,
\end{equation}
where $\boldsymbol{z}^k_m$'s are  Gaussian signal sources with different means and equal variances, i.e., $\boldsymbol{z}_m^k \! \sim \!N(\eta_m^k,\sigma_z^2)$,  $\boldsymbol{\nu}_{k}$'s are Gaussian measurement noises  $\boldsymbol{\nu}_{k}\sim N(0,\sigma_{\nu}^2)$, and $\boldsymbol{z}_m^k,\boldsymbol{\nu}_{k}$ are all mutually uncorrelated. Each sensor applies a local rule to decide which of the $M$ hypotheses has occurred, such that the error probability at the sensor is minimized, i.e., the local detector of sensor $k$   finds $l_k=\arg\min_m |\boldsymbol{x}_k-\eta_m^k|$ and decides hypothesis $\mathcal{H}_{k}$. Let $p_{im}^k$ denote the probability that sensor $k$ decides on $\mathcal{H}_i$, given that the true hypothesis is $\mathcal{H}_m$. For the sensing model in (\ref{sensing-model}), one can verify that $p_{im}^k$ is:
\begin{eqnarray}\label{Pimk}
p_{im}^k=
\begin{cases}
Q(\frac{\eta_i^k+\eta^k_{i-1}-2\eta^k_m}{2(\sigma_{\nu}^2+\sigma_z^2)})-Q(\frac{\eta_i^k+\eta^k_{i+1}-2\eta^k_m}{2(\sigma_{\nu}^2+\sigma_z^2)})  & i\neq 1,M\\
1-Q(\frac{\eta_1^k+\eta^k_{2}-2\eta^k_m}{2(\sigma_{\nu}^2+\sigma_z^2)})  & i=1\\
Q(\frac{\eta^k_M+\eta^k_{M-1}-2\eta^k_m}{2(\sigma_{\nu}^2+\sigma_z^2)})  & i=M
\end{cases}
\end{eqnarray}
where the $Q$-function is defined as 
$Q(x)=\frac{1}{\sqrt{2 \pi}} \int_x^{\infty}e^{-t^2/2}dt$. 
Sensor $k$ employs an $M$-ary digital modulator to map its $M$-ary decision to a symbol and transmits this symbol with power $P_{dk}$.  We consider $M$-PSK and $M$-FSK modulation at the sensors. Let $ \boldsymbol{u}_k$ and $\underline{\boldsymbol{u}}_k$ denote the modulated symbol at sensor $k$ corresponding to $M$-PSK and $M$-FSK modulation, respectively, where scalar 
$ \boldsymbol{u}_k \in \{ e^{j2\pi \frac{i-1}{M}}, i=1,...,M \} $, $M \times 1$ vector
$ \underline{\boldsymbol{u}}_k \in \{ \underline{e}_i , i=1,...,M\} $ and $\underline{e}_i$ is an $M \times 1$ canonical vector whose all elements except the $i$-th one are zeros. We refer to the modulated symbols $\boldsymbol{u}_k$, $\underline{\boldsymbol{u}}_k$ as {\it data symbols} and $P_{dk}$ as {\it data symbol transmit power} corresponding to sensor $k$. Assuming the data symbols are sent over orthogonal channels between sensors and the FC, we represent the channel output at the FC upon the reception of data symbols as a $N \times 1$ vector  $\underline{\boldsymbol{y}}_d=\left[\boldsymbol{y}_{d1}; ...;\boldsymbol{y}_{dk};...;\boldsymbol{y}_{dN} \right]$ for $M$-PSK modulation and $MN \times 1$
 $\underline{\boldsymbol{y}}_d=\left[\underline{\boldsymbol{y}}_{d1} ;...;\underline{\boldsymbol{y}}_{dk};...;\underline{\boldsymbol{y}}_{dN}\right]$ for $M$-FSK modulation. In particular, the channel output corresponding to sensor $k$ is:
\begin{align}\label{system-model}
\boldsymbol{y}_{dk}&=\sqrt{P_{dk}}\boldsymbol{h}_{k}\boldsymbol{u}_k+\boldsymbol{n}_{dk}, ~~~~k=1,...,N~~ M\mbox{-PSK}, \nonumber\\
\underline{\boldsymbol{y}}_{dk}&=\sqrt{P_{dk}}{h}_{k}\underline{\boldsymbol{u}}_k+\underline{\boldsymbol{n}}_{dk}, ~~~~k=1,...,N~~ M\mbox{-FSK}.
\end{align}
The communication channel noises, denoted as scalar $\boldsymbol{n}_{dk}$ and $M \times 1$ vector $\underline{\boldsymbol{n}}_{dk}$, are identically distributed zero mean complex Gaussian $\boldsymbol{n}_{dk}\sim CN(0,\sigma_n^2)$,  $\underline{\boldsymbol{n}}_{dk}\sim CN(0,\sigma_n^2I)$, where $I$ is an $M \times M$ identity matrix. 
We assume that the channel outputs conditioned on the channel inputs, are independent across the sensors. 
The complex channel coefficient $\boldsymbol{h}_k$ in (\ref{system-model})
 is modeled as $\boldsymbol{h}_k\sim CN(0,\sigma^2_{h_k})$ where $\boldsymbol{h}_k=\mathbf{\alpha}_k e^{j\mathbf{\phi}_k}$, the amplitude $\mathbf{\alpha}_k$ and the phase $\mathbf{\phi}_k$, have Rayleigh and uniform distributions, respectively. 
To enable training based channel estimation, we assume that the channel coefficients are fixed for two consecutive symbol intervals, and each sensor sends a training symbol with power $P_{tk}$ along with its data symbol. We refer to the symbols $u_t$, $\underline{u}_t$ as {\it training symbols} and $P_{tk}$ as {\it training symbol transmit power} corresponding to sensor $k$. Without loss of generality, we assume $u_t\!=\!1$ and $\underline{u}_t\!=\!\underline{e}_1$, respectively, when the sensors employ $M$-PSK and $M$-FSK modulation schemes. Training symbols are also sent over orthogonal channels between sensors and the FC, prior to sending data symbols. The channel output corresponding to sensor $k$ at the FC upon the reception of training symbol is:
\begin{align}\label{training-model}
\boldsymbol{y}_{tk}&=\sqrt{P_{tk}}\boldsymbol{h}_{k}u_{t}+\boldsymbol{n}_{tk},  ~~~~k=1,...,N~~ M\mbox{-PSK}, \nonumber\\
\underline{\boldsymbol{y}}_{tk}&=\sqrt{P_{tk}}\boldsymbol{h}_{k}\underline{u}_{t}+\underline{\boldsymbol{n}}_{tk}, ~~~~k=1,...,N~~ M\mbox{-FSK},
\end{align}
where the identically distributed zero mean complex Gaussian noises $\boldsymbol{n}_{tk}\sim CN(0,\sigma_n^2)$,  $\underline{\boldsymbol{n}}_{tk}\sim CN(0,\sigma_n^2 I)$ are independent from $\boldsymbol{n}_{dk}$ and $\underline{\boldsymbol{n}}_{dk}$  in (3). We consider both coherent and non-coherent receivers. 
The unknown channel parameters to be estimated depend on the receiver structure. For a coherent receiver with a training based channel estimator and a non-coherent receiver with a training based channel amplitude estimator, the unknown parameters are $\boldsymbol{h}_k$ and $\boldsymbol{\alpha}_k$, respectively. We model these as 
$\boldsymbol{h}_k=\boldsymbol{\hat{h}}_k+\boldsymbol{\tilde{h}}_k$ and $\boldsymbol{\alpha}_k=\boldsymbol{\hat{\alpha}}_k+\boldsymbol{\tilde{\alpha}}_k$, where $\boldsymbol{\hat{h}}_k$ and $\boldsymbol{\hat{\alpha}}_k$ are the estimates based on $\boldsymbol{y}_{tk}$ and $\boldsymbol{v}_{tk}=|\boldsymbol{y}_{tk}^1|^2$ in (\ref{training-model}) respectively, and $\boldsymbol{\tilde{h}}_k$ and $\boldsymbol{\tilde{\alpha}}_k$ are the estimation errors\footnote{As we mentioned in Section I, the sensors employ $M$-FSK modulation when the FC is equipped with a non-coherent receiver and a training based channel amplitude estimator. Since $\underline{u}_t=\underline{e}_1$, the estimator only employs the first entry of vector $\underline{\boldsymbol{y}}_{tk}$, denoted as $\boldsymbol{y}_{tk}^1$, for channel amplitude estimation.}.
Suppose there is a constraint on the network transmit power, i.e., $\sum_{k=1}^N  P_k\leq P_{tot}$,  where $P_k= P_{dk} + P_{tk}$ is total transmit power for sending data and training symbols at sensor $k$. {\it This modeling allows us to consider the energy cost of channel estimation in network power allocation}. Let $P_d=\sum_{k=1}^{N}P_{dk}$ be the total power devoted for sending all data symbols to FC and, $P_t=\sum_{k=1}^{N}P_{tk}$ be the total power devoted for sending all training symbols to the FC. With this notation, we find $P_d+P_{t}\leq P_{tot}$. We define $r=P_d/P_{tot}$ as the fraction of the power assigned to all data symbols and $r_k=P_{dk}/P_k$  for $k=1,...,N$ as the fraction of the power assigned to the data symbol at sensor $k$ where $r, r_k\in [0,1]$. 
\newline Our goal is to derive J-divergence in closed form expression. For $M$-ary hypothesis testing, \cite{goodman} defined J-divergence as weighted pairwise J-divergence between hypothesis $\mathcal{H}_i$ and $\mathcal{H}_j$, that is $J_{tot|\boldsymbol{\hat{G}}}=(1/2) \sum_{i=1}^M \sum_{j=1}^M \pi_i \pi_j  J_{ij}$ where $J_{ij}= J(f(\boldsymbol{\underline{y}}_d|\mathcal{H}_i,\boldsymbol{\hat{G}}),f(\boldsymbol{\underline{y}}_d|\mathcal{H}_j,\boldsymbol{\hat{G}}))$,  where $\boldsymbol{\hat{G}}\!=diag\lbrace \boldsymbol{\hat{h}}_1, ..., \boldsymbol{\hat{h}}_N\rbrace $ for coherent receiver, $\boldsymbol{\hat{G}}\!=\!diag\left\lbrace \boldsymbol{\hat{\alpha}}_1, ..., \boldsymbol{\hat{\alpha}}_N\right\rbrace $ for non-coherent receiver with a training based channel amplitude estimator, and $\boldsymbol{\hat{G}}$ is null for non-coherent receiver without a channel estimator, in order to incorporate the knowledge of the FC about the channel parameters, and the average J-divergence $J_{avg}=\mathbb{E}_{\boldsymbol{\hat{G}}}\lbrace J_{tot|\boldsymbol{\hat{G}}}\rbrace $.
\section{Optimal fusion rules}
In this section, we adopt the Bayesian criterion to find the optimal fusion rule at the FC, in order to make a global decision $u_0\in \left\lbrace \mathcal{H}_1,\mathcal{H}_2,..,\mathcal{H}_M\right\rbrace $. The optimal fusion rule is $u_0=\arg\max_m \pi_m  \Theta_m$ where $\Theta_m$ varies, depending on the receiver structure and the modulation scheme. Note that the channel outputs are independent across sensors. When sensors employ $M$-PSK and the receiver is coherent we find:
\begin{eqnarray}\label{general-fusion-rule-1}
\!\!\!\!\!\!\!\!\!\Theta_m \!\!&\!\!= \!\!& \!\! \prod_{k=1}^N f(\boldsymbol{y}_{dk}|\mathcal{H}_m) \!= \! \prod_{k=1}^N \sum_{i=1}^M p_{im}^kf(\boldsymbol{y}_{dk}|\boldsymbol{u}_k(i), \boldsymbol{\hat{G}}(k)).
 \end{eqnarray}\label{general-fusion-rule}
When sensors employ $M$-FSK and the receiver is non-coherent we obtain:
\begin{eqnarray}\label{general-fusion-rule-2}
\Theta_m  \!\!&\!\!= \!\!& \!\! \prod_{k=1}^N \sum_{i=1}^M p_{im}^kf(\underline{\boldsymbol{y}}_{dk}|\underline{\boldsymbol{u}}_k(i), \boldsymbol{\hat{G}}(k)), 
\end{eqnarray}
where $\boldsymbol{\hat{G}}(k)$ is the $k$-th diagonal entry of $\boldsymbol{\hat{G}}$, $\boldsymbol{u}_k(i)$, $\underline{\boldsymbol{u}}_k(i)$ in (\ref{general-fusion-rule-1}), (\ref{general-fusion-rule-2}) are the transmitted data symbols of sensor $k$ corresponding to the decision of $\mathcal{H}_i$ and $p_{im}^k$ is obtained from (\ref{Pimk}).
\subsection{Coherent Reception with Imperfect CSI}
For the linear signal model in (\ref{training-model}), the minimum mean square error (MMSE) of channel estimation $\boldsymbol{h}_k $ given $\boldsymbol{y}_{tk}$ is $\boldsymbol{\hat{h}}_k=\mathbb{E}\left\lbrace \boldsymbol{h}_k|\boldsymbol{y}_{tk}\right\rbrace= \frac{\sigma_{h_k}^2 \sqrt{P_{tk}}}{\sigma^2_{h_k} P_{tk}+\sigma_n^2}\boldsymbol{y}_{tk}$. Considering (4), we observe that $\boldsymbol{y}_{tk}\sim CN(0,P_{tk}\sigma^2_{h_k}+\sigma^2_n)$. Let $\gamma_{tk}=P_{tk}/\sigma^2_n$. Since $\boldsymbol{\hat{h}}_k$ is a linear function of $\boldsymbol{y}_{tk}$, we have $\boldsymbol{\hat{h}}_k\sim CN(0, \frac{\sigma^4_{h_k}\gamma_{tk}}{1+\sigma^2_{h_k}\gamma_{tk}})$ and $\boldsymbol{\tilde{h}}_k\sim CN(0,\frac{\sigma^2_{h_k}}{1+\sigma^2_{h_k}\gamma_{tk}})$ \cite{6516873}.
To find $f(\boldsymbol{y}_{dk}|\boldsymbol{u}_k(i),\boldsymbol{\hat{h}}_k)$ in (\ref{general-fusion-rule-1}), we realize that
given $\boldsymbol{u}_k(i)$ and $\boldsymbol{\hat{h}}_k$, we have $\boldsymbol{y}_{dk}\sim CN(\sqrt{P_{dk}}\boldsymbol{\hat{h}}_k\boldsymbol{u}_k(i),\sigma^2_{w_k})$ where $\sigma^2_{w_k}=P_{dk}\sigma^2_{{\tilde{h}}_k}+\sigma^2_n$. Therefore, one can write the conditional pdf $f(\boldsymbol{y}_{dk}|\boldsymbol{u}_k(i),\boldsymbol{\hat{h}}_k)$ and find $\Theta_m$. 
After eliminating  the terms that are independent of $m$, the fusion rule reduces to $u_0=\arg\max_m \pi_m \Theta_m$ where $\Theta_m$ is:
   \begin{equation}\label{Theta-m-coh}
   \!\!\!\!
  \Theta_m= \prod_{k=1}^N \sum_{i=1}^M p_{im}^k \exp\left( \frac{2\sqrt{P_{dk}}Re(e^{\frac{-j2\pi(i-1)}{M}}\boldsymbol{y}_{dk}\boldsymbol{\hat{h}}_k^* )}{\sigma_{w_k}^2}\right). 
   \end{equation}
Note that the optimal fusion rule depends on $P_k, r_k$ (through $P_{dk},\sigma_{w_k}^2$), channel outputs $\boldsymbol{y}_{dk}$, channel estimates $\boldsymbol{\hat{h}}_k$, and local sensor performance indices $p_{im}^k$.  For the special case of $M\!=\!2$, the optimal fusion rule reduces to:
\begin{equation*}
      \!\!\!
\sum_{k=1}^N      {\log\left( \frac{p_{22}^k+(1-p_{22}^k)e^{-\frac{4\sqrt{P_{dk}}}{\sigma^2_{w_k}}Re({\boldsymbol{y}_{dk}}\boldsymbol{\hat{h}}_k^*)}}{p_{21}^k+(1-p_{21}^k)e^{-\frac{4\sqrt{P_{dk}}}{\sigma^2_{w_k}}Re({\boldsymbol{y}_{dk}}\boldsymbol{\hat{h}}_k^*)}}\right) }\overset{H_1}{\underset{H_0}{\gtreqqless}}{\log\left( \frac{\pi_0}{\pi_1}\right)}.
 \end{equation*} 
\subsection{Non-coherent Reception with Imperfect Channel Amplitude}
The MMSE estimate of the channel amplitude $\boldsymbol{\alpha}_k$ given $\boldsymbol{v}_{tk}=|\boldsymbol{y}^1_{tk}|^2$ is $\boldsymbol{\hat{\alpha}}_k=\mathbb{E}\left\lbrace \boldsymbol{\alpha}_k|\boldsymbol{v}_{tk}\right\rbrace =\int \alpha_kf(\alpha_k|\boldsymbol{v}_{tk})d\alpha_k$, where  the conditional pdf $f(\boldsymbol{\alpha}_k|\boldsymbol{v}_{tk})$ is \cite{6516873}: 
      \begin{equation*}\begin{split}
      f(\boldsymbol{\alpha}_k|\boldsymbol{v}_{tk})=&2\boldsymbol{\alpha}_k(1+\gamma_{tk})\exp(\frac{\gamma_{tk} \boldsymbol{v}_{tk}}{(1+\gamma_{tk}) \sigma^2_n}-(1+\gamma_{tk}\boldsymbol{\alpha}^2_k))\\& \times I_0(2\boldsymbol{\alpha}_k\sqrt{\gamma_{tk}\frac{\boldsymbol{v}_{tk}}{\sigma^2_n}}),\\
     \end{split} \end{equation*}
     and $I_0(.)$ is the modified Bessel functin of the first kind with order zero. Given $\boldsymbol{v}_{tk}$, we have $\boldsymbol{\alpha}_k\sim \mathrm{Rice}(\nu,s^2)$ where $\nu=\frac{1}{\gamma_{tk}+1}\sqrt{\gamma_{tk}\frac{\boldsymbol{v}_{tk}}{\sigma^2_n}}$ and $s^2=\frac{1}{\gamma_{tk}+1}$. Therefore, $\boldsymbol{\hat{\alpha}}_k$ is \cite{6516873}:
    \begin{equation*}
     \boldsymbol{\hat{\alpha}}_k=\frac{\sqrt{\pi s^2}}{2}F_1(\frac{-1}{2},1;\frac{-\nu^2}{s^2}),
     \end{equation*}
     where $F_1(.,.;.)$ is the Kummer confluent hyper-geometric function and $F_1(\frac{-1}{2},1;x)=e^{\frac{x}{2}}(xI_1(\frac{x}{2})-(x-1)I_0(\frac{x}{2}))$, $I_1(.)$ is the modified Bessel function of the first kind with order one.
    Furthermore, the variance of estimation error can be computed as below \cite{6516873}:
     \begin{equation}\label{variance error channel}
     \sigma^2_{\tilde{\alpha}_k}=1-\frac{\pi}{4}\frac{1}{\gamma_{tk}+1}\mathbb{E}\left\lbrace F_1(\frac{-1}{2},1;\frac{-\nu^2}{s^2})^2\right\rbrace. 
     \end{equation} 
     Substituting  $\boldsymbol{\hat{\alpha}}_k$ in (3), we have:
        \begin {equation*}
        \underline{\boldsymbol{y}}_{dk} = \sqrt{P_{dk}}\boldsymbol{\hat{\alpha}}_k e^{j\boldsymbol{\phi}_k} \underline{\boldsymbol{u}}_k+\underline{\boldsymbol{w}}_k, ~\mbox{where}~ \underline{\boldsymbol{w}}_k=\sqrt{P_{dk}}\boldsymbol{\tilde{\alpha}}_k  e^{j\boldsymbol{\phi}_k}\underline{\boldsymbol{u}}_k+\underline{\boldsymbol{n}}_{dk}. 
        \end{equation*}
     To find $f(\underline{\boldsymbol{y}}_{dk}|\underline{\boldsymbol{u}}_k(i),\boldsymbol{\hat{\alpha}}_k)$ we write $f(\underline{\boldsymbol{y}}_{dk}|\underline{\boldsymbol{u}}_k(i),\boldsymbol{\hat{\alpha}}_k)=\int f(\underline{\boldsymbol{y}}_{dk}|\underline{\boldsymbol{u}}_k(i),\boldsymbol{\hat{\alpha}}_k, \phi_k) f(\phi_k)d\phi_k$. However, to express $f(\underline{\boldsymbol{y}}_{dk}|\underline{\boldsymbol{u}}_k(i),\boldsymbol{\hat{\alpha}}_k, \phi_k)$ we need the conditional pdf $f(\underline{\boldsymbol{w}}_k|\underline{\boldsymbol{u}}_k(i),\boldsymbol{\hat{\alpha}}_k, \phi_k)$. Unfortunately, this conditional pdf depends on the pdf $f(\boldsymbol{\hat{\alpha}}_k)$ and finding its closed form expression is mathematically intractable. However, our simulation results suggest that, conditional $\underline{\boldsymbol{w}}_k$ can be approximated as a zero-mean complex Gaussian vector with a  covariance matrix $\boldsymbol{C}_{\underline{w}_k}$ whose entries are  $\boldsymbol{C}_{\underline{w}_k}(j,j)=\sigma^2_n$ for  $j\neq i$ and $\boldsymbol{C}_{\underline{w}_k}(j,j)=\sigma^2_{w_k}=P_{dk}\sigma^2_{{\tilde{\alpha}}_k}+\sigma^2_n$ for $ j =i$. Consequently, given $\underline{\boldsymbol{u}}_k(i),\boldsymbol{\hat{\alpha}}_k$ and $\phi_k$, we can approximate $\underline{\boldsymbol{y}}_{dk}\sim CN(\sqrt{P_{dk}}\boldsymbol{\hat{\alpha}}_k e^{j\phi_k} \underline{\boldsymbol{u}}_k(i),\boldsymbol{C}_{w_k})$. With this approximation, we proceed with finding 
     $f(\underline{\boldsymbol{y}}_{dk}|\underline{\boldsymbol{u}}_k(i),\boldsymbol{\hat{\alpha}}_k)$.
     One can verify the following:
     \begin{equation}\begin{aligned}
         &f( \underline{\boldsymbol{y}}_{dk}|\underline{\boldsymbol{u}}_k(i),\boldsymbol{\hat{\alpha}}_k) \\
     &=\frac{c_1c_2(|\boldsymbol{y}_{dk}^i|)}{2\pi}\int_0^{2\pi}
         \exp\left( \frac{2 \sqrt{P_{dk}} Re\left( \boldsymbol{y}_{dk}^i\boldsymbol{\hat{\alpha}}_k e^{-j\phi_k}\right) }{\sigma^2_{w_k}}\right) d\phi_k\\
         &\overset{a}{=}\frac{c_1c_2(|\boldsymbol{y}_{dk}^i|)}{2\pi}\int_0^{2\pi}\exp\left( \frac{2\sqrt{P_{dk}}\boldsymbol{\hat{\alpha}}_k|\boldsymbol{y}_{dk}^i|\cos(\phi_k-\theta)}{\sigma^2_{w_k}}\right) d\phi_k\\
            &=c_1c_2\left(| \boldsymbol{y}_{dk}^i|\right) I_0\left( \frac{2\sqrt{P_{dk}}\boldsymbol{\hat{\alpha}}_k}{\sigma^2_{w_k}}|\boldsymbol{y}_{dk}^i|\right),\\
       \end{aligned} \end{equation}
     in which $c_1\!=\!\frac{\exp{(\frac{-P_{dk}|\boldsymbol{\hat{\alpha}}_k|^2}{\sigma^2_{w_k}})}}{\sqrt{(\pi)^M(\sigma^2_{n})^{M-1}\sigma^2_{w_k}}}{\exp (-\sum_{j=1}^{N}\frac{|\boldsymbol{y}_{dk}^j|^2}{\sigma^2_{n}})}$,
        $c_2(|\boldsymbol{y}_{dk}^i|)\!=\!\exp(\frac{P_{dk}\sigma^2_{{\tilde{\alpha}}_k}|\boldsymbol{y}_{dk}^i|^2}{\sigma^2_n\sigma^2_{w_k}})$. To obtain (a), we let $\boldsymbol{y}_{dk}^i=|\boldsymbol{y}_{dk}^i|e^{j\theta}$. After substituting $f(\underline{\boldsymbol{y}}_{dk}|\underline{\boldsymbol{u}}_k(i),\boldsymbol{\hat{\alpha}}_k)$ in (\ref{general-fusion-rule-2}) and eliminating $c_1$ due to its irrelevance to $m$, the optimal fusion rule reduces to $u_0=\arg\max_m\pi_m \Theta_m$ where $\Theta_m$ is:
        \begin{equation}\label{Theta-FSK-training}
       \Theta_m= \prod_{k=1}^{N}\sum_{i=1}^{M}p_{im}^k\underbrace{c_2\left( |\boldsymbol{y}_{dk}^i|\right) I_0\left( \frac{2\sqrt{P_{dk}}\boldsymbol{\hat{\alpha}}_k}{\sigma^2_{w_k}}|\boldsymbol{y}_{dk}^i|\right) }_{=F(|\boldsymbol{y}_{dk}^i|)}.
        \end{equation}
     Note that the optimal fusion rule depends on $P_k, r_k$ (through $P_{dk},\sigma_{w_k}^2$), magnitude of channel outputs $|\boldsymbol{y}_{dk}^i|$, channel amplitude estimates $\boldsymbol{\hat{\alpha}}_k$, and local sensor performance indices $p_{im}^k$.
     For the  special case of $M\!=\!2$ the optimal fusion rule reduces to:
           \begin{equation}\label{binary-coh}
           \sum_{k=1}^{N} \log\left( \frac{(1-p_{22}^k)F(\boldsymbol{y}^1_{dk})+p_{22}^kF(\boldsymbol{y}^2_{dk})}{(1-p_{21}^k)F(\boldsymbol{y}^1_{dk})+p_{21}^k F(\boldsymbol{y}^2_{dk})}\right) \overset{H_1}{\underset{H_0}{\gtreqless}} \log\left( \frac{\pi_0}{\pi_1}\right). 
           \end{equation}
  %
     \subsection{Non-coherent Reception with Channel Statistics}
     In the absence of training, we have $P_{dk}=P_{k}$ and $P_{tk}=0$.  To find $f(\underline{\boldsymbol{y}}_{dk}|\underline{\boldsymbol{u}}_k(i))$ in (\ref{general-fusion-rule-2}), we 
     realize that given $\underline{\boldsymbol{u}}_k(i)$, we have $\underline{\boldsymbol{y}}_{dk}\sim CN(0,\boldsymbol{C}_{u_k(i)})$ where $\boldsymbol{C}_{u_k(i)}$ is a diagonal matrix whose entries are $\boldsymbol{C}_{u_k(i)}(j,j)\!=\!\sigma^2_n$ for $j \! \neq \! i$ and $\boldsymbol{C}_{u_k(i)}(j,j)\!=\!P_{dk} \sigma_{h_k}^2 + \sigma^2_n$ for $j \!= \!i$. We can verify that $f(\underline{\boldsymbol{y}}_{dk}|\underline{\boldsymbol{u}}_k(i))$ equals to:
      \begin{equation*}
        \beta
         \exp\left( \frac{ P_{dk}\sigma_{h_k}^2|\boldsymbol{y}_{dk}^i|^2}{\sigma^2_n\left( \sigma^2_n+P_{dk} \sigma_{h_k}^2\right) }\right)\prod_{j=1}^{M} \exp\left( -\frac{|\boldsymbol{y}_{dk}^j|^2}{\sigma^2_n}\right). 
     \end{equation*}
     After substituting $f(\underline{\boldsymbol{y}}_{dk}|\underline{\boldsymbol{u}}_k(i))$ in (\ref{general-fusion-rule-2}) and eliminating \\ $\prod_{j=1}^{M}\exp(-\frac{|\boldsymbol{y}_{dk}^j|^2}{\sigma^2_n})$ due to its irrelevance to $m$, the optimal fusion rule reduces to $u_0=\arg\max_m\pi_m \Theta_m$ where $\Theta_m$ is:
     \begin{equation}\label{theta-non-training}
     \Theta_m=\prod_{k=1}^{N}
         \sum_{i=1}^{M}p_{im}^k \underbrace{\exp\left( \frac{P_{dk}\sigma_{h_k}^2|\boldsymbol{y}_{dk}^i|^2}{\sigma^2_n\left(  \sigma^2_n+P_{dk} \sigma_{h_k} ^2\right)  }\right) }_{=H(|\boldsymbol{y}_{dk}^i|)}.
         \end{equation}
     Different from (\ref{Theta-FSK-training}), (\ref{theta-non-training}) does not depend on channel amplitude estimates and only depends on the channel statistics. For the special case of $M=2$, the optimal fusion rule is similar to (\ref{binary-coh}) with the difference that  $F(\boldsymbol{y}_{dk}^i)$ needs to be replaced with $H(|\boldsymbol{y}_{dk}^i|)$ defined in (\ref{theta-non-training}).
     
\section{Optimal Power allocations}
In this section, we drive J-divergence for different receivers. Then, we optimize power allocation between sensors based on J-divergence.
\subsection{Coherent Reception with Imperfect CSI}
 In order to compute conditional J-divergence for coherent reception, we need the distribution of $f( \boldsymbol{\underline{y}}_d|\boldsymbol{\hat{G}},\mathcal{H}_i )$ and $f( \boldsymbol{\underline{y}}_d|\boldsymbol{\hat{G}},\mathcal{H}_j )$. The pdf of the received signals $\boldsymbol{\underline{y}}_d$ at the FC given the transmitted signals $\boldsymbol{\underline{u}}=[\boldsymbol{u}_1;...;\boldsymbol{u}_N]$ from the sensors is:
  \begin{equation}\label{conditional distribution}
   \!\!\!f\left( \boldsymbol{\underline{y}}_d| \boldsymbol{\underline{u}}, \hat{\boldsymbol{G}}\right)\!=\!\frac{1}{|2\pi\boldsymbol{R}|^{\frac{1}{2}}}\exp\left[ -\frac{1}{2}\left(\boldsymbol{\underline{y}}_d-\hat{\boldsymbol{G}}{A}\boldsymbol{\underline{u}}\right)^T\!\!\boldsymbol{R}^{-1}\!\!\left(\boldsymbol{\underline{y}}_d-\hat{\boldsymbol{G}}{A}\boldsymbol{\underline{u}}\right)\right],    
   \end{equation}
   where $A=diag\left\lbrace \sqrt{P_{d1}},...,\sqrt{P_{dN}}\right\rbrace $ and $\boldsymbol{R}=diag\left\lbrace\sigma^2_{w_1},...,\sigma^2_{w_N}\right\rbrace $ are diagonal matrices.
 The conditional pdf of $\boldsymbol{\underline{y}}_d$ given the two hypotheses are:
  \begin{equation}\label{markov}
    f\left( \boldsymbol{\underline{y}}_d|\hat{\boldsymbol{G}},\mathcal{H}_i \right) =\sum_{\boldsymbol{\underline{u}}}f\left( \boldsymbol{\underline{u}}|\mathcal{H}_i\right)f\left( \boldsymbol{\underline{y}}_d| \hat{\boldsymbol{G}},\boldsymbol{\underline{u}}\right).     
   \end{equation}
 As we can see, Both  distributions $f( \boldsymbol{\underline{y}}_d|\hat{\boldsymbol{G}},\mathcal{H}_i )$ and $f( \boldsymbol{\underline{y}}_d|\hat{\boldsymbol{G}},\mathcal{H}_j )$ have Gaussian mixture distributions. Unfortunately, the J-divergence between
two Gaussian mixture densities does not have a general
closed-form expression \cite{vincentpoor}. We observe that when $ \mathbb{E}\left\lbrace Tr\left(A^T\hat{\boldsymbol{G}}^T\boldsymbol{R}^{-1}\hat{\boldsymbol{G}}A\right)\right\rbrace\rightarrow 0$ and $ \mathbb{E}\left\lbrace Tr\left(A^T\hat{\boldsymbol{G}}^T\boldsymbol{R}^{-1}\hat{\boldsymbol{G}}A\right)^2\right\rbrace\rightarrow 0$ , $f( \boldsymbol{\underline{y}}_d|\hat{\boldsymbol{G}}, \mathcal{H}_i )$ can be approximated by a Gaussian distribution with mean vector $\mu_i$ and covariance matrix ${\Sigma}_i$. We find $\mu_i$ and $\boldsymbol{\Sigma}_i$ using moment matching Markov properties. 
\begin{eqnarray}
\mu_i &=& \int_{\boldsymbol{\underline{y}}_d}\boldsymbol{\underline{y}}_d f( \boldsymbol{\underline{y}}_d|\hat{\boldsymbol{G}}, \mathcal{H}_i )d\boldsymbol{\underline{y}}_d\\
&=&\int_{\boldsymbol{\underline{y}}_d}\boldsymbol{\underline{y}}_d\sum_{\boldsymbol{\underline{u}}}f\left( \boldsymbol{\underline{y}}_d| \hat{\boldsymbol{G}}, \boldsymbol{\underline{u}}\right) f\left( \boldsymbol{\underline{u}}|\mathcal{H}_i\right)d\boldsymbol{\underline{y}}_d\\
&=& \sum_{\boldsymbol{\underline{u}}}f\left( \boldsymbol{\underline{u}}|\mathcal{H}_i\right)\int_{\boldsymbol{\underline{y}}_d}\boldsymbol{\underline{y}}_d f\left( \boldsymbol{\underline{y}}_d| \boldsymbol{\underline{u}}\right)d\boldsymbol{\underline{y}}_d
\end{eqnarray}
Recall that $f( \boldsymbol{\underline{y}}_d| \boldsymbol{\underline{u}}, \hat{\boldsymbol{G}})$ is a Gaussian density with mean $\hat{\boldsymbol{G}}{A}\boldsymbol{\underline{u}}$, as shown in (\ref{conditional distribution}), hence:
\begin{eqnarray}\label{mean of y}
\mu_i&=&\sum_{\boldsymbol{\underline{u}}}\hat{\boldsymbol{G}}{A}\boldsymbol{\underline{u}}f\left( \boldsymbol{\underline{u}}|\mathcal{H}_i\right)= \hat{\boldsymbol{G}}{A}\underline{\beta}_i
\end{eqnarray}
Where $\underline{\beta}_i$ is a $1 \times N$  vector and equals to: 
\begin{equation}
\begin{split}
\underline{\beta}_i&=\sum_{\boldsymbol{\underline{u}}}\boldsymbol{\underline{u}}f( \boldsymbol{\underline{u}}|\mathcal{H}_i)\\
\underline{\beta}_i(k)&=\sum_{l=1}^M e^{j2\pi\frac{l-1}{M}}p_{li}^k,\ k=1,..., N\\
\end{split}
\end{equation}
Similarly, we can compute $\boldsymbol{\Sigma}_i$ as below:
\allowdisplaybreaks
\begin{eqnarray}\label{covariance of y}
{\Sigma}_i=\int_{\boldsymbol{\underline{y}}_d}\left[ \boldsymbol{\underline{y}}_d-\mu_i\right]\left[ \boldsymbol{\underline{y}}_d-\mu_i\right]^Tf\left( \boldsymbol{\underline{y}}_d|\hat{\boldsymbol{G}},\mathcal{H}_i\right)\nonumber\\
=\left[ \boldsymbol{\underline{y}}_d-\mu_i\right]\left[ \boldsymbol{\underline{y}}_d-\mu_i\right]^T\sum_{\boldsymbol{\underline{u}}}f\left( \boldsymbol{\underline{y}}_d|\hat{\boldsymbol{G}},\boldsymbol{\underline{u}}\right)f\left( \boldsymbol{\underline{u}}|\mathcal{H}_i\right)\nonumber\\
=\sum_{\boldsymbol{\underline{u}}}f\left( \boldsymbol{\underline{u}}|\mathcal{H}_i\right)\int_{\boldsymbol{\underline{y}}_d}\left[ \boldsymbol{\underline{y}}_d-\hat{\boldsymbol{G}}A\boldsymbol{\underline{u}}+\hat{\boldsymbol{G}}A\boldsymbol{\underline{u}}-\mu_i\right]\nonumber\\ 
\times \left[ \boldsymbol{\underline{y}}_d-\hat{\boldsymbol{G}}A\boldsymbol{\underline{u}}+\hat{\boldsymbol{G}}A\boldsymbol{\underline{u}}-\mu_i\right]^T f( \boldsymbol{\underline{y}}_d|\boldsymbol{\underline{u}})d\boldsymbol{\underline{y}}_d\nonumber\\
= \boldsymbol{R}+\sum_{\boldsymbol{\underline{u}}}f\left( \boldsymbol{\underline{u}}|\mathcal{H}_i\right)\left[\hat{\boldsymbol{G}}A\boldsymbol{\underline{u}}-\mu_i\right]\left[\hat{\boldsymbol{G}}A\boldsymbol{\underline{u}}-\mu_i\right]^T. 
\end{eqnarray}
The last step follows because $f(\boldsymbol{\underline{y}}_d|\hat{\boldsymbol{G}},\boldsymbol{\underline{u}})$ is a Gaussian density with mean $\hat{\boldsymbol{G}}A\boldsymbol{\underline{u}}$ and covariance matrix $\boldsymbol{R}$. Applying (\ref{mean of y}), and after some algebra, we obtain:
\allowdisplaybreaks
\begin{equation}
\boldsymbol{\Sigma}_i=\boldsymbol{R}+\hat{\boldsymbol{G}}AB_iA^T\hat{\boldsymbol{G}}^T,
\end{equation}
where the diagonal matrix $B_i$ is the covariance matrix of $f( \boldsymbol{\underline{u}}|\mathcal{H}_i)$ and equals 
$B_i\!\!\!=\!\!\!cov(\boldsymbol{\underline{u}}|\mathcal{H}_i)=diag\lbrace cov(\boldsymbol{u}_1|\mathcal{H}_i), ..., cov(\boldsymbol{u}_N|\mathcal{H}_i)\rbrace$ and $cov(\boldsymbol{u}_k|\mathcal{H}_i)=1-\underline{\beta}_i(k)$. Next, based on J-divergence between two Guassian distribution, $J_{ij|\hat{\boldsymbol{G}}}$ becomes:
\allowdisplaybreaks
\begin{align*}
J_{ij|\hat{\boldsymbol{G}}}&=\frac{1}{2}Tr\left[\boldsymbol{\Sigma}_i\boldsymbol{\Sigma}_j^{-1}+ \boldsymbol{\Sigma}_j\boldsymbol{\Sigma}_i^{-1}\right. \\
&\left. \left( \boldsymbol{\Sigma}_i^{-1}+\boldsymbol{\Sigma}_j^{-1}\right) \left( \mu_i-\mu_j\right)\left( \mu_i-\mu_j\right)^T\right] -N. 
\end{align*}
Using  $\mu_i$ and $\boldsymbol{\Sigma}_i$ in (\ref{mean of y}) and (\ref{covariance of y}), and after some algebra, $J_{tot|\hat{\boldsymbol{G}}}$ divergence will become:
\begin{eqnarray*}
\label{conditional_j}
\!\!\!\!\!\!\!J_{tot|\hat{\boldsymbol{G}}}&\!\!=\!\!&\frac{1}{2}\sum_{i=1}^{M}\sum_{j=1}^{M}\sum_{k=1}^{N}\frac{\hat{\boldsymbol{g}}_kP_{dk}\gamma_{ji}^k+\sigma^2_{w_k}}{\hat{\boldsymbol{g}}_kP_{dk}B_i(k)+\sigma^2_{w_k}}\\
&&+\frac{\hat{\boldsymbol{g}}_kP_{dk}\gamma_{ij}^k+\sigma^2_{w_k}}{\hat{\boldsymbol{g}}_kP_{dk}B_j(k)+\sigma^2_{w_k}}-N,
\end{eqnarray*}
where $\gamma_{ij}^k=B_i(k)+|\mu_i(k)-\mu_j(k)|^2$ and $\gamma_{ji}^k=B_j(k)+|\mu_i(k)-\mu_j(k)|^2$ and $\hat{\boldsymbol{g}}_k=|\hat{\boldsymbol{h}}_k|^2$.
The average J-divergence for coherent reception has been calculated in (\ref{avg_jdiv}).
\begin{figure*}[!t]
\begin{align}\label{avg_jdiv}
\!\!\!\!\!J_{avg}&=\mathbb{E}\left( J_{tot|\hat{\boldsymbol{G}}}\right)  =\int\frac{1}{2}\sum_{i=1}^{M}\sum_{j=1}^{M}\sum_{k=1}^{N}\frac{\hat{\boldsymbol{g}}_kP_{dk}\gamma_{ji}^k+\sigma^2_{w_k}}{\hat{\boldsymbol{g}}_kP_{dk}B_i(k)+\sigma^2_{w_k}}
+\frac{\hat{\boldsymbol{g}}_kP_{dk}\gamma_{ij}^k+\sigma^2_{w_k}}{\hat{\boldsymbol{g}}_kP_{dk}B_j(k)+\sigma^2_{w_k}}f\left(\hat{\boldsymbol{g}}_k \right) d{\hat{\boldsymbol{g}}}_k-N \nonumber\\
& = \sum_{i=1}^{M}\sum_{j=1}^{M}\sum_{k=1}^{N}\frac{\gamma_{ji}^k}{B_i(k)}+\frac{\gamma_{ij}^k}{B_j(k)}
+\int\frac{1-{\frac{\gamma_{ji}^k}{B_i(k)}}}{\frac{\hat{\boldsymbol{g}}_kP_{dk}\gamma_{ji}^k}{\sigma^2_{w_k}}+1}+\frac{1-{\frac{\gamma_{ij}^k}{B_j(k)}}}{\frac{\hat{\boldsymbol{g}}_kP_{dk}\gamma_{ij}^k}{\sigma^2_{w_k}}+1}f(\hat{\boldsymbol{g}}_k)d\hat{\boldsymbol{g}}_k-N\\
& = \sum_{i=1}^{M}\sum_{j=1}^{M}\sum_{k=1}^{N}\frac{\gamma_{ji}^k}{B_i(k)}+\left(1-\frac{\gamma_{ji}^k}{B_i(k)} \right) \frac{x_k}{B_i(k)}e^{\frac{x_k}{B_i(k)}}E_1\left(\frac{x_k}{B_i(k)} \right)+\frac{\gamma_{ij}^k}{B_j(k)}+ \left(1-\frac{\gamma_{ij}^k}{B_j(k)} \right)\frac{x_k}{B_j(k)}e^{\frac{x_k}{B_j(k)}}E_1\left(\frac{x_k}{B_j(k)}\right)-N\nonumber
\end{align}
\end{figure*}
In (\ref{avg_jdiv}) $x_k$ is:
\begin{equation*}
x_k=\frac{1}{r_k(1-r_k)}(\frac{1}{\sigma^2_k/\sigma^2_nP_k}+(\frac{1}{\sigma^2_k/\sigma^2_nP_k})^2)
\end{equation*}
$E_1(x)=\int_{x}^{\infty}\frac{e^{-t}}{t}dt$ is the exponential integral.
Defining $s_k=\frac{\sigma^2_{h_k}P_k}{\sigma^2_n}$ as the recieved SNR of sensor k at the FC, We can see that $x_k$ is a function of recieved SNR's of sensors at the fusion center. 
Let $D(x)=xe^xE_1(x)$, then average J-divergence will reduce to:
\newline
\setlength{\abovedisplayskip}{3pt}
\setlength{\belowdisplayskip}{3pt}
\begin{eqnarray*}\label{average j divergence}
J_{avg}&=&C+\sum_{k=1}^{N}\sum_{i=1}^{M}\sum_{j=1}^{M}\left(1-\frac{\gamma_{ji}^k}{B_i(k)}\right) D\left( \frac{x_k}{B_i(k)}\right) \\
&&+\left(1-\frac{\gamma_{ij}^k}{B_j(k)}\right) D\left( \frac{x_k}{B_j(k)}\right) 
\end{eqnarray*}
Where $C$ is:
\begin{equation*}
C=\frac{1}{2}\sum_{i=1}^{M}\sum_{j=1}^{M}\sum_{k=1}^{N}\frac{\gamma_{ji}^k}{B_i(k)}+\frac{\gamma_{ij}^k}{B_j(k)}-N.
\end{equation*}
\textbf{Optimal power allocation based on $J_{tot|\hat{\boldsymbol{G}}}$}:
The formulation of optimization problem will be as below:
\begin{eqnarray}\label{optimization_tot}
&& ({\boldsymbol{\cal P}}_1): \max_{\left(P_{d1},P_{t1}\right),...,\left(P_{dN},P_{tN}\right)} J_{tot|\hat{\boldsymbol{G}}}(P_{d1},...,P_{dN})\\
&& \mbox{s.t.}  \sum_{k=1}^{N}P_{dk} \leq P_{d}, \sum_{k=1}^{N}P_{tk} \leq P_{t}, P_{dk}, P_{tk} \geq 0, P_{d}+P_{t} \leq P_{tot}\nonumber
\end{eqnarray}
As we can see from our cost function,The objective function is fully decoupled, a direct result of the orthogonal channels
between the sensors and the FC.
\newline Lemma 1: The first order derivative of the objective function
$J_{tot|\hat{\boldsymbol{G}}}(P_{1}, · · · , P_{dN})$ with respect to $P_{dj}$ and $P_{tj}$ is always nonnegative at
any valid power allocation point $P_{dk} \geq 0$. That is
\begin{equation}
\begin{split}
\frac{\partial}{\partial P_{dj}}
J_{tot|\hat{\boldsymbol{G}}}(P_{d1}, · · · , P_{dN})|_{P_{dk}\geq 0}\geq 0\\
\frac{\partial}{\partial P_{tj}}
J_{tot|\hat{\boldsymbol{G}}}(P_{t1}, · · · , P_{tN})|_{P_{dk}\geq 0}\geq 0.
\end{split}
\end{equation}
This lemma has been proved in Appendix \ref{sec:lemma1}.
\newline Lemma 1 tells us that the objective function in (\ref{optimization_tot}) is
nondecreasing with increasing power budget $P_{tot}$. Since we
are maximizing a nondecreasing function, the optimal point
is always at the constraint boundary, i.e.,$\sum_{k=1}^{N}P_{dk}+P_{tk}=P_{tot}$. This result is intuitively
plausible since it makes full use of the power budget. However, its not obvious what portion of total power should be devoted to $P_d$ and the remaining to $P_t$. In this case, we assume we know $r$. Therefore, the optimization will be broken to two different parts: one for determining $P_{dj}$ and the other part for determining $P_{tj}$.  
\newline For Data power allocation, the optimization problem reduces to the solution to the following optimization problem,
\begin{equation}\begin{split}
\label{opt_coh_j}
\max_{P_{d1},...,P_{dN}}& J_{tot| \hat{\boldsymbol{G}}}(P_{d1},...,P_{dN})\\
s.t &\sum_{k=1}^{N}P_{dk}\leq P_{d},\  P_{dk}\geq 0\\
\end{split}\end{equation}
\newline Similar to lemma 1, we can prove the second order derivative of the objective function
$J_{tot|\hat{\boldsymbol{G}}}(P_{1}, · · · , P_{dN})$ with respect to $P_{dk}$ is always non positive at
any valid power allocation point $P_{dk} \geq 0$. That is
\begin{equation}
\frac{\partial^2}{\partial P_{dj}^2}
J_{tot|\hat{\boldsymbol{G}}}|_{P_{dj}\geq 0} \leq 0.
\end{equation}
The Lagrangian associated with
the constrained optimization problem in (\ref{opt_coh_j}) is 
\begin{align}
& L(P_{d1},...,P_{dN})=\nonumber\\
&\frac{1}{2}\sum_{i=1}^{M}\sum_{j=1}^{M}\sum_{k=1}^{N}\frac{\hat{\boldsymbol{g}}_kP_{dk}\gamma_{ji}^k+\sigma^2_{w_k}}{\hat{\boldsymbol{g}}_kP_{dk}B_i(k)+\sigma^2_{w_k}}
+\frac{\hat{\boldsymbol{g}}_kP_{dk}\gamma_{ij}^k+\sigma^2_{w_k}}{\hat{\boldsymbol{g}}_kP_{dk}B_j(k)+\sigma^2_{w_k}}\nonumber\\
&   -\lambda(\sum_{k=1}^{N}P_{dk}-P_{d})+\sum_{k=1}^{N}\nu_{dk}P_{dk}\nonumber
\end{align}
There is no closed form solution for $P_{dk}$'s and the optimal values can be obtained through general convex optimization techniques. However, its obvious that $P_{dj}^*$ is a function of performance of sensors, channel estimation coefficients and channel estimation errors. 
\newline For training power allocation, the optimization problem reduces to the solution to the following optimization problem,
\begin{equation}\begin{split}
\max_{P_{t1},...,P_{tN}}& J_{tot| \hat{\boldsymbol{G}}}(P_{t1},...,P_{tN})\\
s.t &\sum_{k=1}^{N}P_{tk}\leq P_{t},\  P_{tk}\geq 0\\
\end{split}\end{equation}
This cost function is a function of $P_{dk}^*$. Therefore , $P_{tk}^*$ is a function of $P_{dk}^*$. As a result, we can not determine $P_{tk}^*$ from conditional J-divergence and we distribute training power uniformly between sensors.
\newline \textbf{Power allocation based on $\boldsymbol{J}_{avg}$}: In the previous section, we optimized data power between sensors based on $J_{tot|\hat{\boldsymbol{G}}}$. We could not derive theoretically $r$ based on $J_{tot|\hat{\boldsymbol{G}}}$ cost function and training power allocation since $J_{tot|\hat{\boldsymbol{G}}}$ is a function of $\hat{\boldsymbol{G}}$. So, we need a cost function to determine how distribute data and training power between sensors. Therefore, we choose $J_{avg}$ for power allocation.
The power allocation for coherent reception based on average J-divergence reduces to the solution to the following optimization problem,
\begin{equation}\begin{split}
\max_{P_{1},...,P_{k},r_{1},...,r_{k}}& J_{avg}(P_{1},...,P_{k},r_{1},...,r_{k}), \text{given in (\ref{conditional_j})}\\
s.t &\sum_{k=1}^{N}P_{k}\leq P_{tot},\ P_{k}\geq 0,\ 0\leq r_{k} \leqslant 1\\
\end{split}\end{equation}
In order to solve this optimization problem, at first we assume we know each $P_k$ and obtain optimal $r_k$.
\newline Lemma 2: If we allocate each sensor power $P_k$ for sending training and data symbol, The optimal power for training and data based on maximizing $J_{avg}$ is $P_{tk}=P_{dk}=P_k/2$. Subsequently, we can conclude $r=r_k=\frac{1}{2}$.
\newline This lemma has been proved in Appendix \ref{sec:lemma2}.This lemma tell us if we want to maximize $J_{avg}$, each sensor should equally distribute the power between training and data or $r=r_k=\frac{1}{2}$.
In lemma 2, we determined optimal $r_k$. In the next step, we determine optimal $P_k$.
\newline Lemma 3: The first order derivative of the objective function
$J_{avg}(P_{1}, ..., P_{N},\frac{1}{2}, ..., \frac{1}{2})$ with respect to $P_{k}$ is always nonnegative at
any valid power allocation point $P_{k} \geq 0$. That is
\begin{equation}
\frac{\partial}{\partial P_{k}}
J_{avg}(P_{1}, ..., P_{N})|_{P_{i}\geq 0}\geq 0.
\end{equation}
\newline\newline This lemma has been proved in Appendix \ref{sec:lemma3}.
From this lemma, we can conclude that $\sum_{k=1}^{N}P_k=P_{tot}$. The second order derivative of $J_{avg}$ is not always negative. Therefore, $J_{avg}$ is not a concave function. So, We only obtain the local maximum for these function through general constrained optimization techniques such as interior point method.

\subsection{Non Coherent Reception with Imperfect Channel Amplitude}
For the non coherent reception, if we want to compute $J_{tot|\boldsymbol{\hat{G}}}$, we need to derive J-divergence between two distributions  $f( \boldsymbol{\underline{y}}_d|\hat{\boldsymbol{G}},\mathcal{H}_i )$ and $f( \boldsymbol{\underline{y}}_d|\hat{\boldsymbol{G}},\mathcal{H}_j )$. Unfortunately there is not a closed form solution for the distance between these distributions. Thus, we will derive $J_{tot|\boldsymbol{\hat{G}},\Phi}$ where $\Phi=diag\left\lbrace \phi_1,..., \phi_N\right\rbrace $.
In order to compute conditional J-divergence for non coherent reception, we need to compute distance between the distribution of $f( \boldsymbol{\underline{y}}_d|\hat{\boldsymbol{G}},{\Phi}, \mathcal{H}_i )$ and $f( \boldsymbol{\underline{y}}_d|\hat{\boldsymbol{G}}, \Phi, \mathcal{H}_j )$.
The conditional density function of the received signals $\boldsymbol{\underline{y}}_d$ at the FC given the transmitted signals $\boldsymbol{\underline{u}}$ from the sensors and $\hat{\boldsymbol{G}},{\Phi}$ is
  \begin{eqnarray}\label{conditional_distributionnon}
   &f\left( \boldsymbol{\underline{y}}_d| \boldsymbol{\underline{u}}, \boldsymbol{\hat{G}}, \Phi\right)=\frac{1}{|2\pi{\boldsymbol{C}_{\underline{w}}}|^{\frac{1}{2}}}\nonumber\\
   & exp\left[ -\frac{1}{2}\left(\boldsymbol{\underline{y}}_d-\Phi\boldsymbol{\hat{G}}{A}\boldsymbol{\underline{u}}\right)^T{\boldsymbol{C}_{\underline{w}}}^{-1}\left(\boldsymbol{\underline{y}}_d-\Phi\boldsymbol{\hat{G}}{A}\boldsymbol{\underline{u}}\right)\right]    
   \end{eqnarray}
   where ${\boldsymbol{C}_{\underline{w}}}=diag\left\lbrace C_{w1},..., C_{wN}\right\rbrace  $.
 The conditional density functions of the received signals given the two hypotheses are
  \begin{equation}\label{markov}
    f\left( \boldsymbol{\underline{y}}_d|\boldsymbol{\hat{G}}, \Phi, \mathcal{H}_i \right) =\sum_{\boldsymbol{\underline{u}}}f\left( \boldsymbol{\underline{y}}_d|\boldsymbol{\hat{G}}, \Phi, \boldsymbol{\underline{u}}\right) f\left( \boldsymbol{\underline{u}}|\mathcal{H}_i\right)    
   \end{equation}
 As we can see similar to coherent reception, Both $f( \boldsymbol{\underline{y}}_d|\boldsymbol{\hat{G}}, \Phi,\mathcal{H}_i )$ and $f( \boldsymbol{\underline{y}}_d|\boldsymbol{\hat{G}}, \Phi,\mathcal{H}_j )$ are Gaussian mixture distributions. Unfortunately, the J-divergence between
two Gaussian mixture densities does not have a general
closed-form expression \cite{vincentpoor}. Similar to coherent reception, when $\mathbb{E}\left\lbrace Tr\left(A^T\hat{\boldsymbol{G}}^T{\boldsymbol{C}_{\underline{w}}}^{-1}\hat{\boldsymbol{G}}A\right)\right\rbrace\rightarrow 0$ and $\mathbb{E}\left\lbrace Tr\left(A^T\hat{\boldsymbol{G}}^T{\boldsymbol{C}_{\underline{w}}}^{-1}\hat{\boldsymbol{G}}A\right)^2\right\rbrace\rightarrow 0$, the conditional distribution function $f( \boldsymbol{\underline{y}}_d|\boldsymbol{\hat{G}}, \mathcal{H}_i )$ can be approximated by a Gaussian distribution with mean vector $\mu_i$ and covariance matrix $\boldsymbol{\Sigma}_i$. We find $\mu_i$ and $\boldsymbol{\Sigma}_i$ using moment matching Markov properties. 
\begin{align}
\underline{\mu}_i&= \int_{\boldsymbol{\underline{y}}_d}\boldsymbol{\underline{y}}_df( \boldsymbol{\underline{y}}_d|\boldsymbol{\hat{G}}, \Phi, \mathcal{H}_i )d\boldsymbol{\underline{y}}_d\nonumber\\
&=\int_{\boldsymbol{\underline{y}}_d}\boldsymbol{\underline{y}}_d\sum_{\boldsymbol{\underline{u}}}f\left( \boldsymbol{\underline{y}}_d| \boldsymbol{\hat{G}}, \Phi, \boldsymbol{\underline{u}}\right) f\left( \boldsymbol{\underline{u}}|\mathcal{H}_i\right)d\boldsymbol{\underline{y}}_d\\
&= \sum_{\boldsymbol{\underline{u}}}f\left( \boldsymbol{\underline{u}}|\mathcal{H}_i\right)\int_{\boldsymbol{\underline{y}}_d}\boldsymbol{\underline{y}}_df\left( \boldsymbol{\underline{y}}_d|\boldsymbol{\hat{G}}, \Phi, \boldsymbol{\underline{u}}\right)d\boldsymbol{\underline{y}}_d\nonumber
\end{align}
Recall that $f\left( \boldsymbol{\underline{y}}_d|\hat{\boldsymbol{G}}, \Phi, \boldsymbol{\underline{u}}\right)$ is a Gaussian density with mean $\hat{\boldsymbol{G}} \Phi{A}\boldsymbol{\underline{u}}$, as shown in (\ref{conditional_distributionnon}), so
\begin{eqnarray}\label{mean of y2}
\underline{\mu}_i=\sum_{\boldsymbol{\underline{u}}} \Phi \hat{\boldsymbol{G}} {A} \boldsymbol{\underline{u}} f\left( \boldsymbol{\underline{u}} | \mathcal{H}_i \right) \nonumber
=\Phi \hat{\boldsymbol{G}} {A} \underline{\beta}_i
\end{eqnarray}
Where $\underline{\beta}_i=\sum_{\boldsymbol{\underline{u}}}\boldsymbol{\underline{u}}f( \boldsymbol{\underline{u}}|\mathcal{H}_i)=\left[\underline{\beta}_i(1),..., \underline{\beta}_i(N)\right]$ is a $1 \times MN$ and $\underline{\beta}_i(k)$ equals
\begin{equation}\begin{split}
\underline{\beta}_i(k)&=\sum_{\boldsymbol{\underline{u}}_k}\boldsymbol{\underline{u}}_kf( \boldsymbol{\underline{u}}_k|\mathcal{H}_i)\\
&=\left[ p_{1i}^k, ..., p_{Mi}^k\right], k=1,...,N\\ 
\end{split}\end{equation}
Similarly, we can compute $\boldsymbol{\Sigma}_i$
\begin{eqnarray}\label{covariance of y2}
\boldsymbol {\Sigma}_i &=& \int_{\boldsymbol{\underline{y}}_d} \left[ \boldsymbol{\underline{y}}_d-\mu_i\right] \left[ \boldsymbol{\underline{y}}_d-\mu_i\right]^T f\left( \boldsymbol{\underline{y}}_d|\mathcal{H}_i\right) d\boldsymbol{\underline{y}}_d \nonumber\\
&=& \left[ \boldsymbol{\underline{y}}_d-\mu_i\right] \left[ \boldsymbol{\underline{y}}_d-\mu_i\right]^T \sum_{\boldsymbol{\underline{u}}}f\left( \boldsymbol{\underline{y}}_d|\boldsymbol{\underline{u}}\right)f\left( \boldsymbol{\underline{u}}|\mathcal{H}_i\right)d\boldsymbol{\underline{y}}_d\nonumber\\
&=&\sum_{\boldsymbol{\underline{u}}}f\left( \boldsymbol{\underline{u}}|\mathcal{H}_i\right)\int_{\boldsymbol{\underline{y}}_d}\left[ \boldsymbol{\underline{y}}_d-\Phi\boldsymbol{\hat{G}}A\boldsymbol{\underline{u}}+\Phi\boldsymbol{\hat{G}}A\boldsymbol{\underline{u}}-\mu_i\right]\nonumber\\ 
&&.\left[ \boldsymbol{\underline{y}}_d-\Phi\boldsymbol{\hat{G}}A\boldsymbol{\underline{u}}+\Phi\boldsymbol{\hat{G}}A\boldsymbol{\underline{u}}-\mu_i\right]^Tf\left( \boldsymbol{\underline{y}}_d|\boldsymbol{\underline{u}}\right)d\boldsymbol{\underline{y}}_d,\nonumber\\
&=& {\boldsymbol{C}_{\underline{w}}}+\sum_{\boldsymbol{\underline{u}}}f\left( \boldsymbol{\underline{u}}|\mathcal{H}_i\right)\left[\Phi\boldsymbol{\hat{G}}A\boldsymbol{\underline{u}}-\mu_i\right]\left[\Phi\boldsymbol{\hat{G}}A\boldsymbol{\underline{u}}-\mu_i\right]^T.\nonumber 
\end{eqnarray}
The last step follows because $f(\boldsymbol{\underline{y}}_d|\hat{\boldsymbol{G}},\boldsymbol{\underline{u}})$ is a Gaussian density with mean $\Phi\boldsymbol{\hat{G}}A\boldsymbol{\underline{u}}$ and covariance matrix ${\boldsymbol{C}_{\underline{w}}}$. Applying (\ref{mean of y2}), and after some algebra, we obtain:
\begin{equation}
\boldsymbol{\Sigma}_i={\boldsymbol{C}_{\underline{w}}}+\boldsymbol{\hat{G}}AB_iA^TD^T,
\end{equation}
where $B_i\!\!\!=cov(\boldsymbol{\underline{u}}|\mathcal{H}_i)=diag\lbrace cov(\boldsymbol{\underline{u}}_1|\mathcal{H}_i),..., cov(\boldsymbol{\underline{u}}_N|\mathcal{H}_i)\rbrace$ is a $NM \times NM$ diagonal matrix and $cov(\boldsymbol{\underline{u}}_k|\mathcal{H}_i)$ is an $M\times M$ diagonal matrix and its $lth$ diagonal element is $p_{li}^k(1-p_{li}^k)$, $k=1,..,N, l=0,...,M-1$.
\newline Next, based on J-divergence between two Guassian distribution, $J_{ij|\hat{\boldsymbol{G}},\Phi}$ will become:
\begin{align*}
J_{ij|\hat{\boldsymbol{G}},\Phi}&=\frac{1}{2}Tr\left[\boldsymbol{\Sigma}_i\boldsymbol{\Sigma}_j^{-1}+ \boldsymbol{\Sigma}_j\boldsymbol{\Sigma}_i^{-1}\right. \\
&\left. \left( \boldsymbol{\Sigma}_i^{-1}+\boldsymbol{\Sigma}_j^{-1}\right) \left( \mu_i-\mu_j\right)\left( \mu_i-\mu_j\right)^T\right] -N.
\end{align*}
Using  $\mu_i$ and $\boldsymbol{\Sigma}_i$ in (\ref{mean of y2}) and (\ref{covariance of y2}), and after some algebra, $J_{tot|\hat{\boldsymbol{G}},\Phi}$ will become:
\begin{eqnarray*}
\!\!\!\!\!\!\!J_{tot|\hat{\boldsymbol{G}}}&\!\!=\!\!&\frac{1}{2}\sum_{i=1}^{M}\sum_{j=1}^{M}\sum_{k=1}^{N}\frac{\hat{\boldsymbol{g}}_kP_{dk}\gamma_{ji}^k+\sigma^2_{w_k}}{\hat{\boldsymbol{g}}_kP_{dk}B_i(k)+\sigma^2_{w_k}}\\
&&+\frac{\hat{\boldsymbol{g}}_kP_{dk}\gamma_{ij}^k+\sigma^2_{w_k}}{\hat{\boldsymbol{g}}_kP_{dk}B_j(k)+\sigma^2_{w_k}}-N,
\end{eqnarray*}
%
Where $\hat{\boldsymbol{g}}_k=\hat{\boldsymbol{\alpha}}_k^2$. As we can see, the conditional J-divergence for non coherent receiver is the same as coherent reception and it does not depend on phase of channel coefficients.
\newline \textbf{Optimal power allocation based on conditional J-divergence}:
As we can see, conditional J-divergence for non coherent reception with amplitude estimation is similar to coherent reception. Thus, data power allocation of non coherent reception is similar to coherent reception. Similarly, for training power allocation between sensors we allocate training power uniformly between sensors. 
\subsection{Non-Coherent Reception with Channel Statistics}
In order to compute total J-divergence for non coherent reception without training, we need the distribution of $f( \boldsymbol{\underline{y}}_d|\mathcal{H}_i )$ and $f( \boldsymbol{\underline{y}}_d|\mathcal{H}_j )$. The pdf of the received signals $\boldsymbol{\underline{y}}_d$ at the FC given the transmitted signals $\boldsymbol{\underline{u}}=[\boldsymbol{\underline{u}}_1,...,\boldsymbol{\underline{u}}_N]$ from the sensors is
  \begin{equation}\label{conditional distribution}
   \!\!\!f( \boldsymbol{\underline{y}}_d| \boldsymbol{\underline{u}})\!=\!\frac{1}{|2\pi\boldsymbol{C}_{\underline{u}}|^{\frac{1}{2}}}\exp\left[ -\frac{1}{2}\boldsymbol{\underline{y}}_d^T \boldsymbol{C}_{\underline{u}}^{-1}\boldsymbol{\underline{y}}_d\right],    
   \end{equation}
where $\boldsymbol{C}_{\underline{u}}=diag\lbrace \boldsymbol{C}_{u_1},..., \boldsymbol{C}_{u_N}\rbrace$.
The conditional pdf of $\boldsymbol{\underline{y}}_d$ given the hypothesise is:
  \begin{equation}\label{markov}
    f( \boldsymbol{\underline{y}}_d|\mathcal{H}_i) =\sum_{\boldsymbol{\underline{u}}}f\left( \boldsymbol{\underline{u}}|\mathcal{H}_i\right)f( \boldsymbol{\underline{y}}_d|\boldsymbol{\underline{u}}).    
   \end{equation}
 As we can see, distribution $f( \boldsymbol{\underline{y}}_d|\mathcal{H}_i )$ has Gaussian mixture distribution.Approximating this Gaussian mixture with a Gaussian distribution, $f(\underline{\boldsymbol{y}}_{d}|\mathcal{H}_i)\sim CN(0,\boldsymbol{\Sigma}_i)$ where $\boldsymbol{\Sigma}_i$ is:
        
        \begin{eqnarray*}
        \boldsymbol{\Sigma}_i&=&\int_{\boldsymbol{\underline{y}}_d} \boldsymbol{\underline{y}}_d \boldsymbol{\underline{y}}_d^Tf\left( \boldsymbol{\underline{y}}_d|\mathcal{H}_i\right)d\boldsymbol{\underline{y}}_d\\
        &=& \sum_{\boldsymbol{\underline{u}}} f\left( \boldsymbol{\underline{u}}|\mathcal{H}_i\right) \int_{\boldsymbol{\underline{y}}_d} \boldsymbol{\underline{y}}_d
         \boldsymbol{\underline{y}}_d^Tf\left( \boldsymbol{\underline{y}}_d|\boldsymbol{\underline{u}}\right)d\boldsymbol{\underline{y}}_d\\
         &=& \sum_{\boldsymbol{\underline{u}}} f\left( \boldsymbol{\underline{u}}|\mathcal{H}_i\right) \boldsymbol{C}_{\underline{u}}\\
         &=& diag\lbrace \sum_{\boldsymbol{\underline{u}_1}} f\left( \boldsymbol{\underline{u}_1}|\mathcal{H}_i\right) \boldsymbol{C}_{\underline{u}_1}, ..., \sum_{\boldsymbol{\underline{u}_N}}f\left( \boldsymbol{\underline{u}_N}|\mathcal{H}_i\right)\boldsymbol{C}_{\underline{u}_N} \rbrace
        \end{eqnarray*}
        where $\sum_{\boldsymbol{\underline{u}}}f\left( \boldsymbol{\underline{u}}|\mathcal{H}_i\right)\boldsymbol{C}_{\underline{u}}$ is a  $M \times M$ diagonal matrix and its $l$th element is $\sigma^2_n+p^k_{li}P_{dk}\sigma^2_{h_k}$.After replacing the distance between two Gaussian distribution in total J-divergence, the total J-divergence will become:
        \begin{eqnarray}
        J_{tot} \!= \!\sum_{i,j,k,l=1}^{M} \pi_{i}\pi_{j}\left[\frac{\sigma^2_n+p^k_{li}P_{dk}\sigma^2_{h_k}}{\sigma^2_n+p^k_{lj}P_{dk}\sigma^2_{h_k}}
        +\frac{\sigma^2_n+p^k_{li}P_{dk}\sigma^2_{h_k}}{\sigma^2_n+p^k_{lj}P_{dk}\sigma^2_{h_k}}\right].
        \end{eqnarray}
       \textbf{Power allocation based on total J-divergence}: 
        The first order derivative of $J_{tot}$ respect to $P_{dk}$ is positive which means $\sum_{k=1}^{N}P_{dk}=P_d$. The second order derivative of these cost functions respect to $P_{dk}$ is positive. Thus, we are maximizing a convex function. We use the following theorem for computation of optimal power allocation based on our cost function.
       \newline Theorem: Let $S$ be a non empty compact polyhedral set on $R^n$ and let $f:S\rightarrow R$ be convex on $S$. An optimal solution $x*$ to the problem $max_{x\in{S}} f(x)$ exists where $x^*$ is an extreme point of $S$ \cite{Mokhtar}.
       \newline Based on above theorem, The optimal power $(P^*_{d1},...,P^*_{dN})$ is an extreme point. So, we obtain all extreme points and choose the one that maximize our cost functions as the optimal point.
       
\section{Numerical results}
        \begin{figure*}[t!]\label{fig:far}
        \centering
        \begin{minipage}[htp]{0.4\textwidth}
          \centering
          \includegraphics[width=1\textwidth]{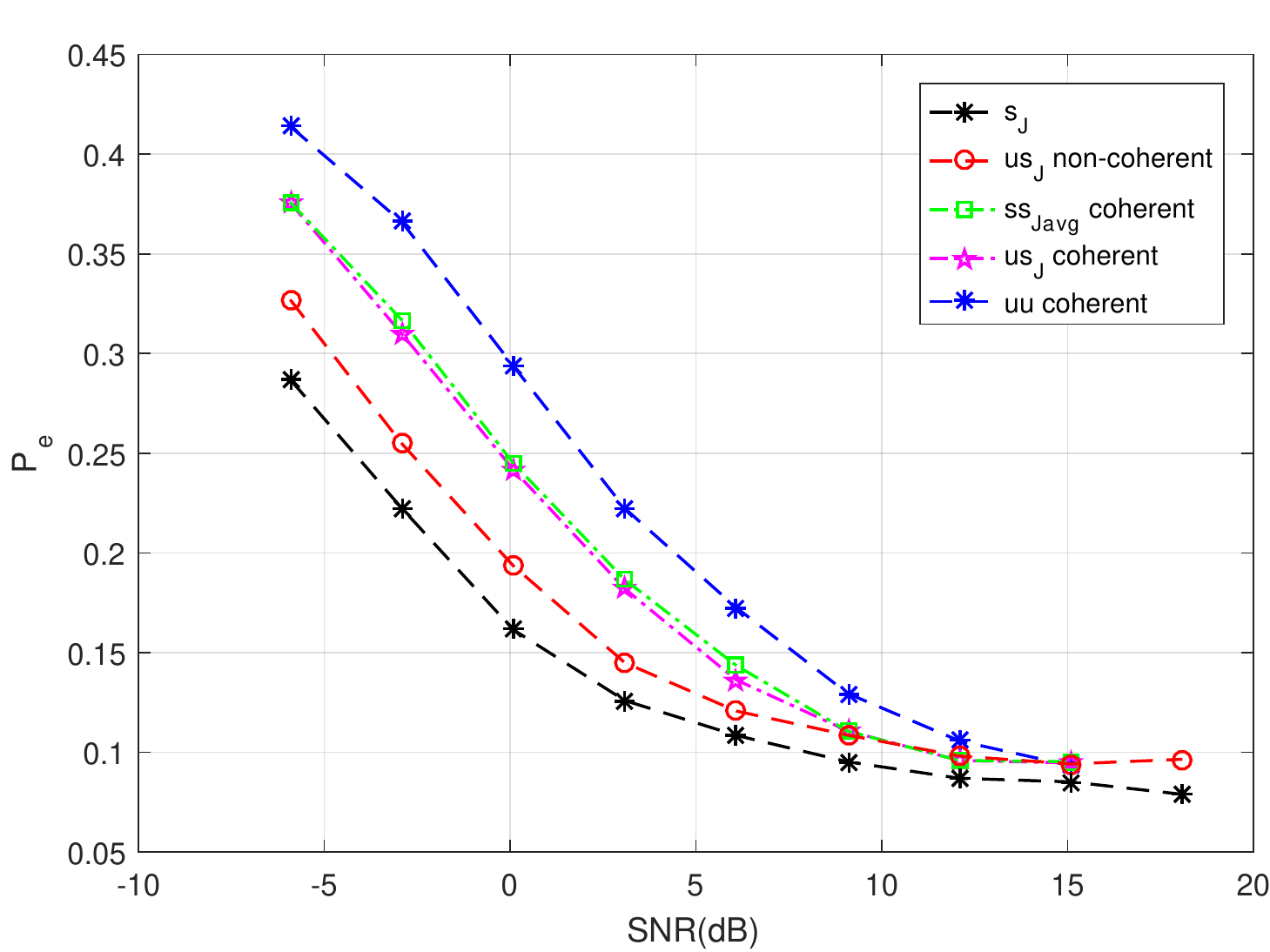}
          \vspace{-0.6cm}
          \centerline{(i)}\medskip
        \vspace{0.4cm}
        \end{minipage}
         \centering  
        \begin{minipage}[htp]{0.4\textwidth}
          \centering
          \includegraphics[width=1\textwidth]{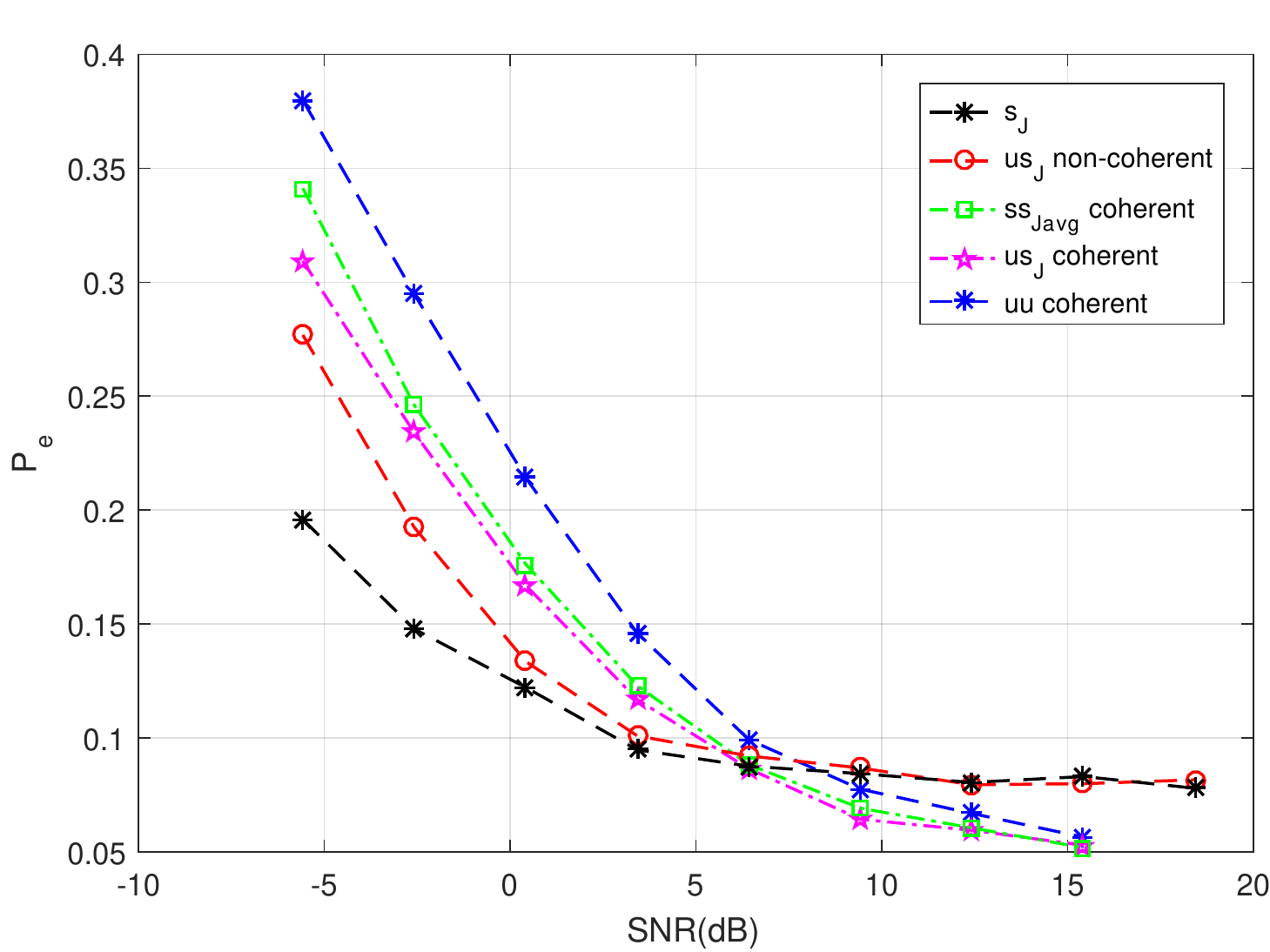}
          \vspace{-0.6cm}
          \centerline{(ii)}\medskip
        \vspace{0.4cm}
        \end{minipage}
         \centering
        \begin{minipage}[htp]{0.4\textwidth}
          \centering
          \includegraphics[width=1\textwidth]{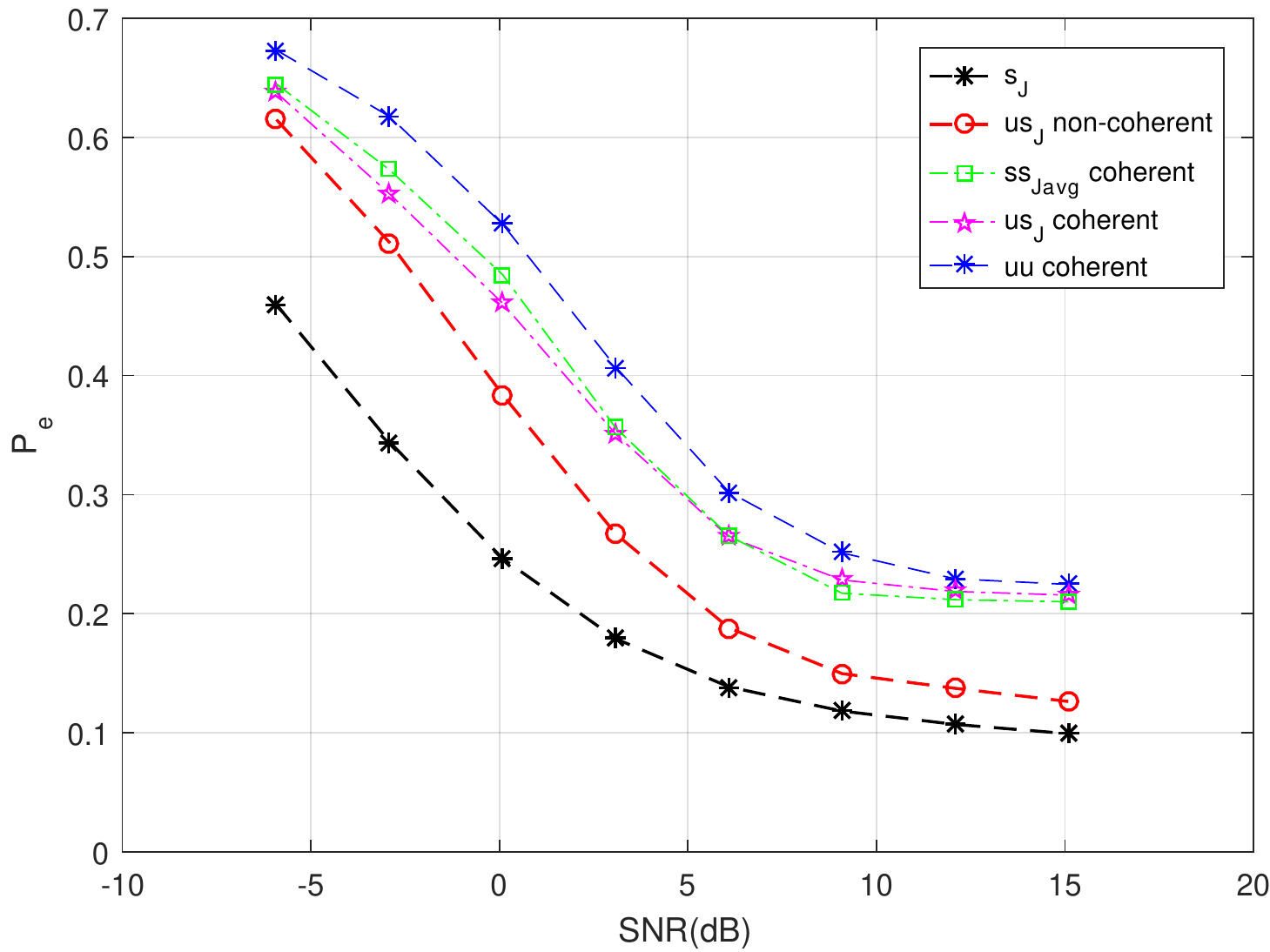}
          \vspace{-0.6cm}
          \centerline{(iii)}\medskip
        \vspace{0.4cm}
        \end{minipage}
          \centering
        \begin{minipage}[htp]{0.4\textwidth}
          \centering
          \includegraphics[width=1\textwidth]{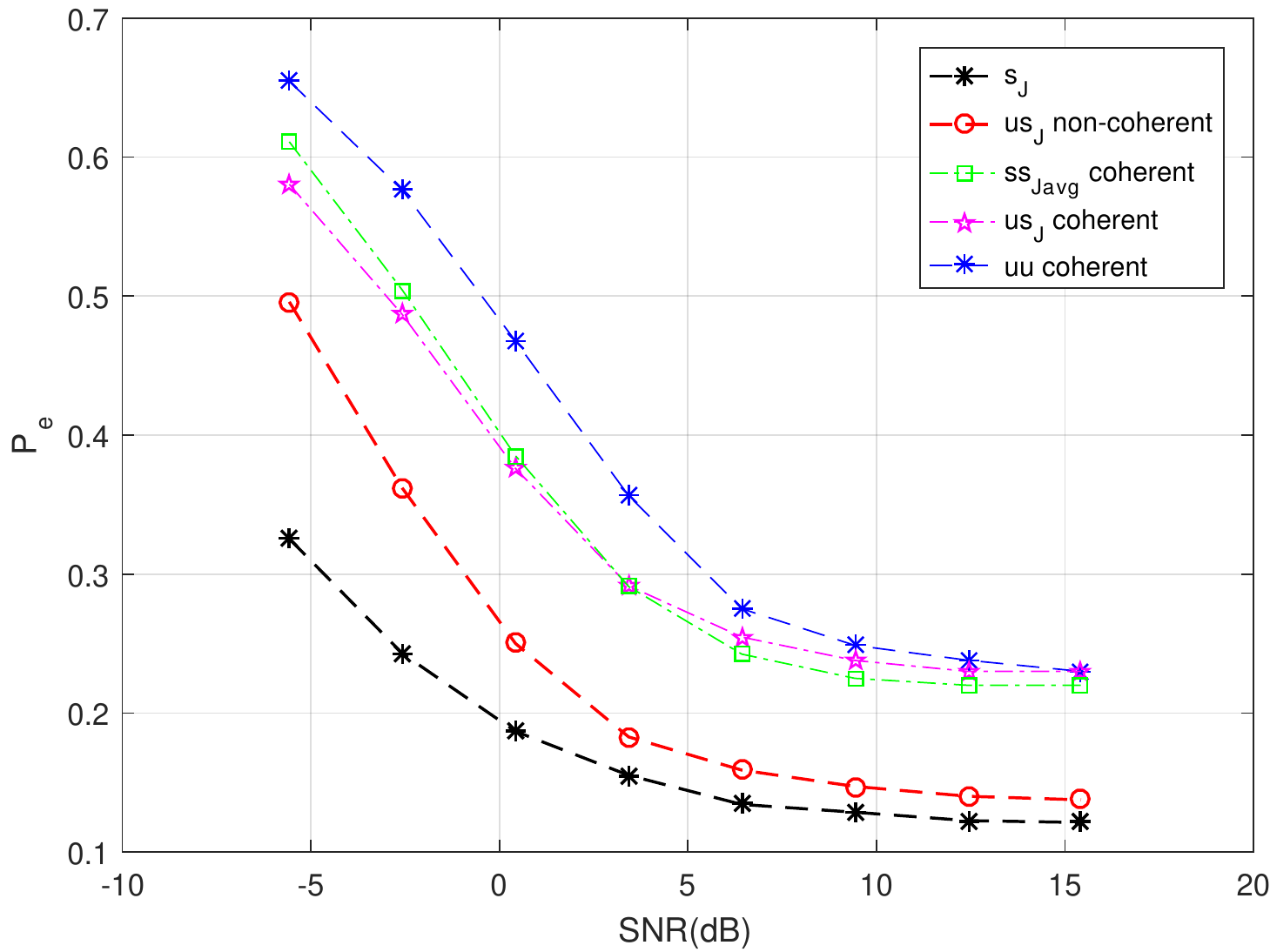}
          \vspace{-0.6cm}
          \centerline{(iv)}\medskip
          \label{fig:x2}
        \vspace{0.4cm}
        \end{minipage}
                \caption{Probability of error vs SNR when sensors closer to FC have less probability of error (i) M=2, N=5, (ii) M=2, N=10, iii) M=4, N=5, iv) M=4, N=10.}
                \end{figure*}
          
        \begin{figure*}[t!]\label{fig:far}
        \centering
        \begin{minipage}[htp]{0.4\textwidth}
          \centering
          \includegraphics[width=1\textwidth]{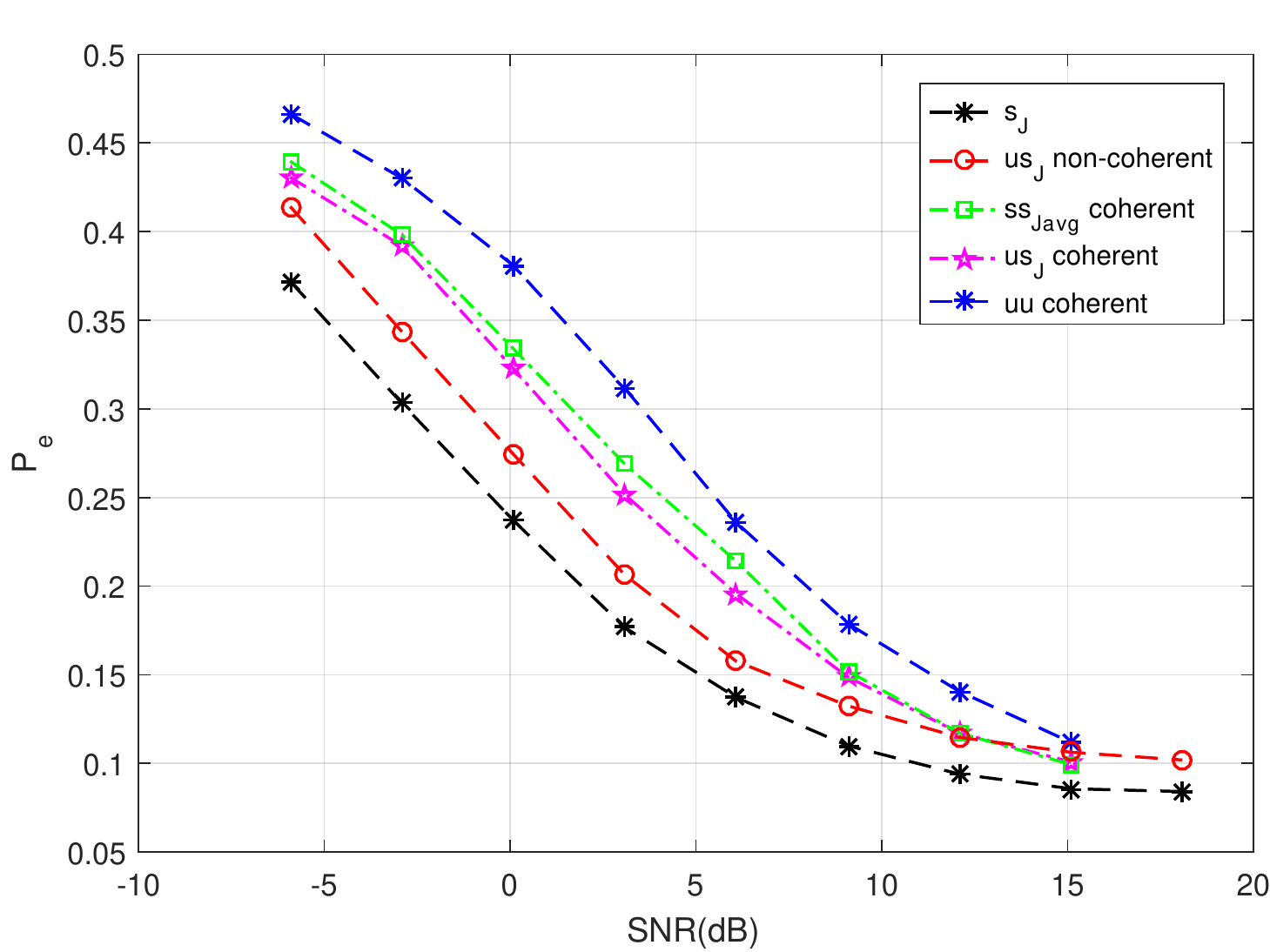}
          \vspace{-0.6cm}
          \centerline{(i)}\medskip
        \vspace{0.4cm}
        \end{minipage}
         \centering  
        \begin{minipage}[htp]{0.4\textwidth}
          \centering
          \includegraphics[width=1\textwidth]{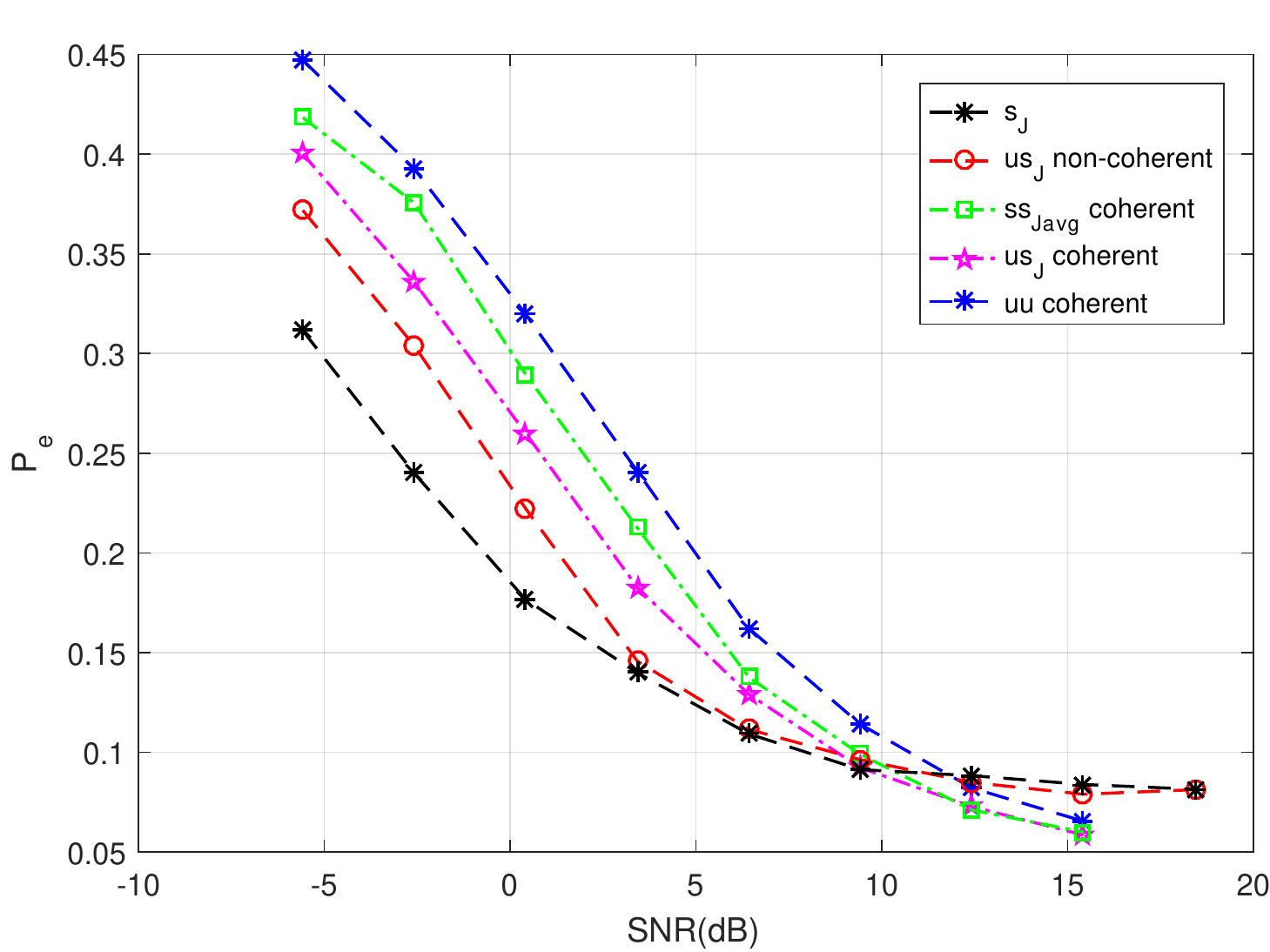}
          \vspace{-0.6cm}
          \centerline{(ii)}\medskip
        \vspace{0.4cm}
        \end{minipage}
         \centering
        \begin{minipage}[htp]{0.4\textwidth}
          \centering
          \includegraphics[width=1\textwidth]{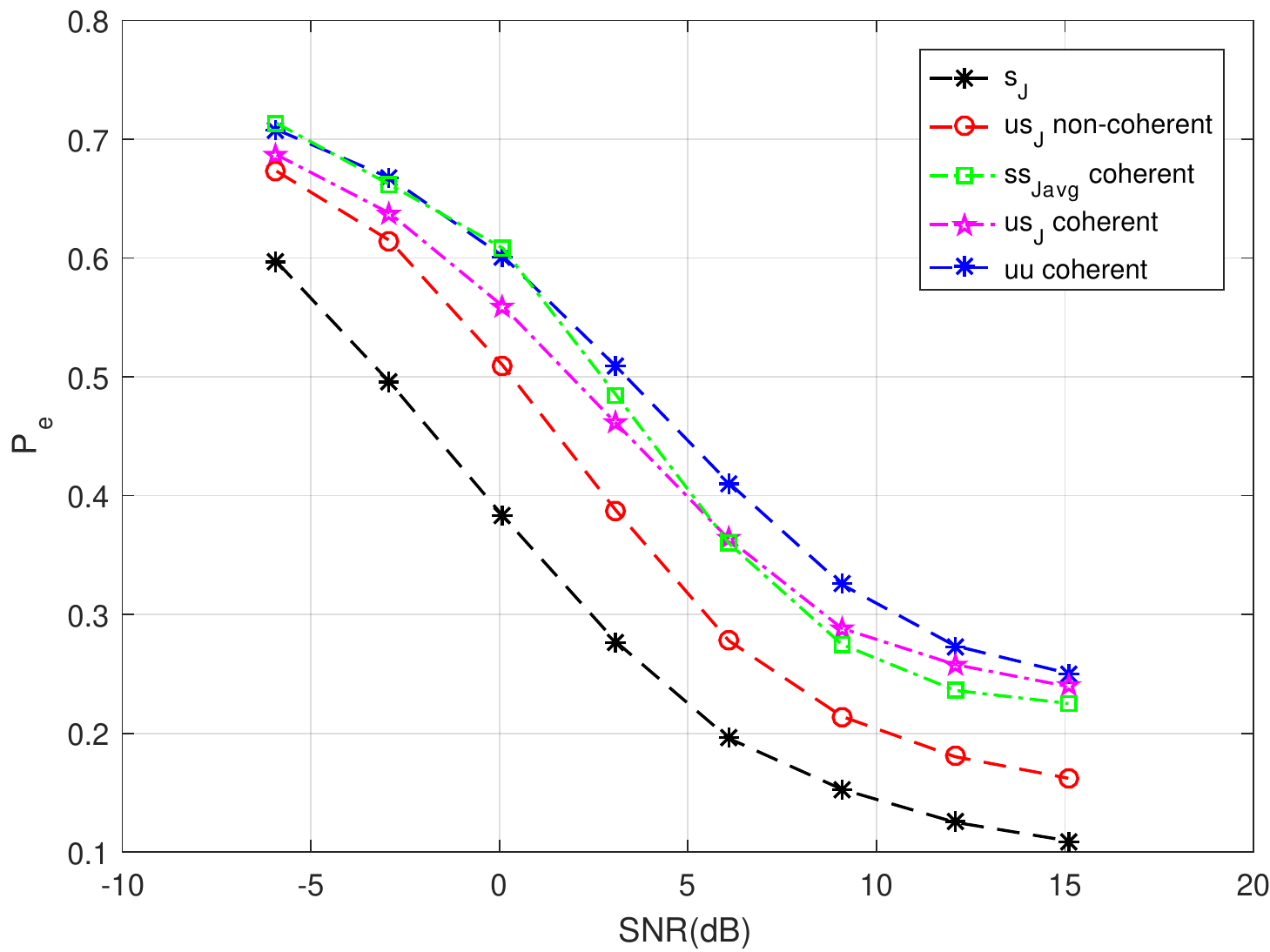}
          \vspace{-0.6cm}
          \centerline{(iii)}\medskip
        \vspace{0.4cm}
        \end{minipage}
          \centering
        \begin{minipage}[htp]{0.4\textwidth}
          \centering
          \includegraphics[width=1\textwidth]{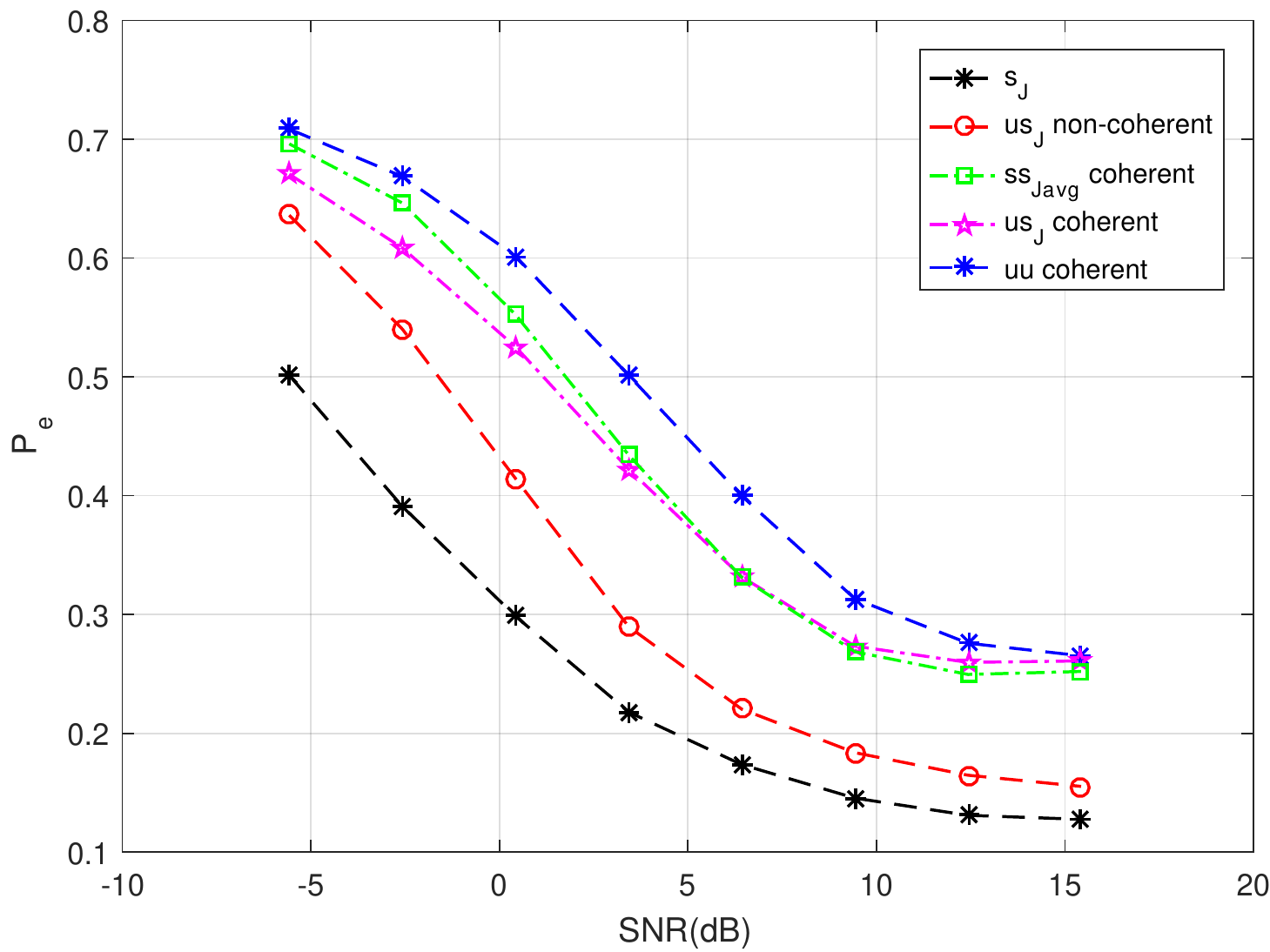}
          \vspace{-0.6cm}
          \centerline{(iv)}\medskip
          \label{fig:x2}
        \vspace{0.4cm}
        \end{minipage}
                \caption{Probability of error vs SNR when sensors farther from FC have less probability of error (i) M=2, N=5, (ii) M=2, N=10, iii) M=4, N=5, iv) M=4, N=10.}
                \end{figure*}
                \begin{figure*}[t!]\label{fig:eqdist}
        \centering
        \begin{minipage}[htp]{0.45\textwidth}
          \centering
          \includegraphics[width=1\textwidth]{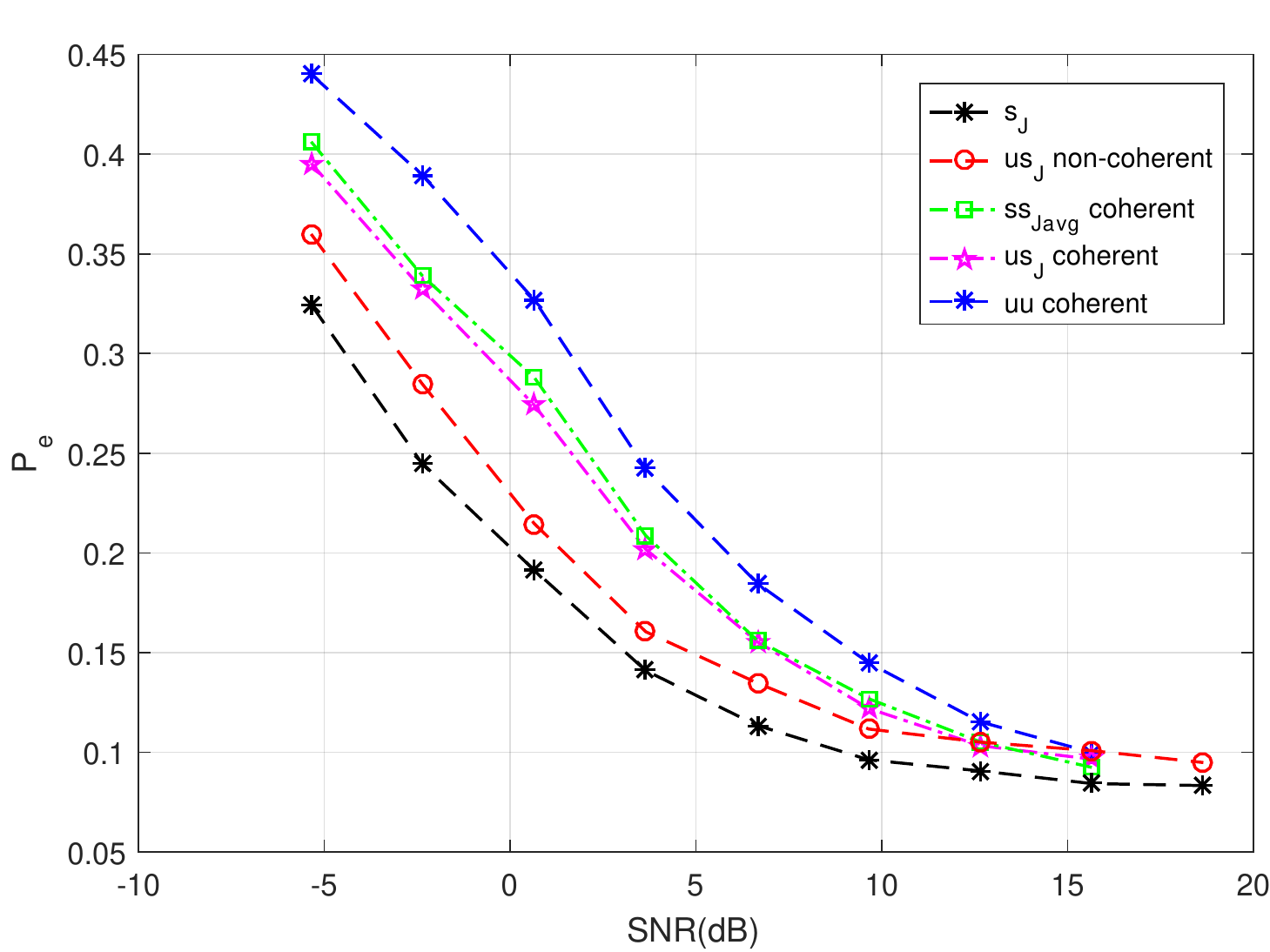}
          \vspace{-0.6cm}
          \centerline{(i)}\medskip
        \vspace{0.4cm}
        \end{minipage}
          \centering
        \begin{minipage}[htp]{0.45\textwidth}
          \centering
          \includegraphics[width=1\textwidth]{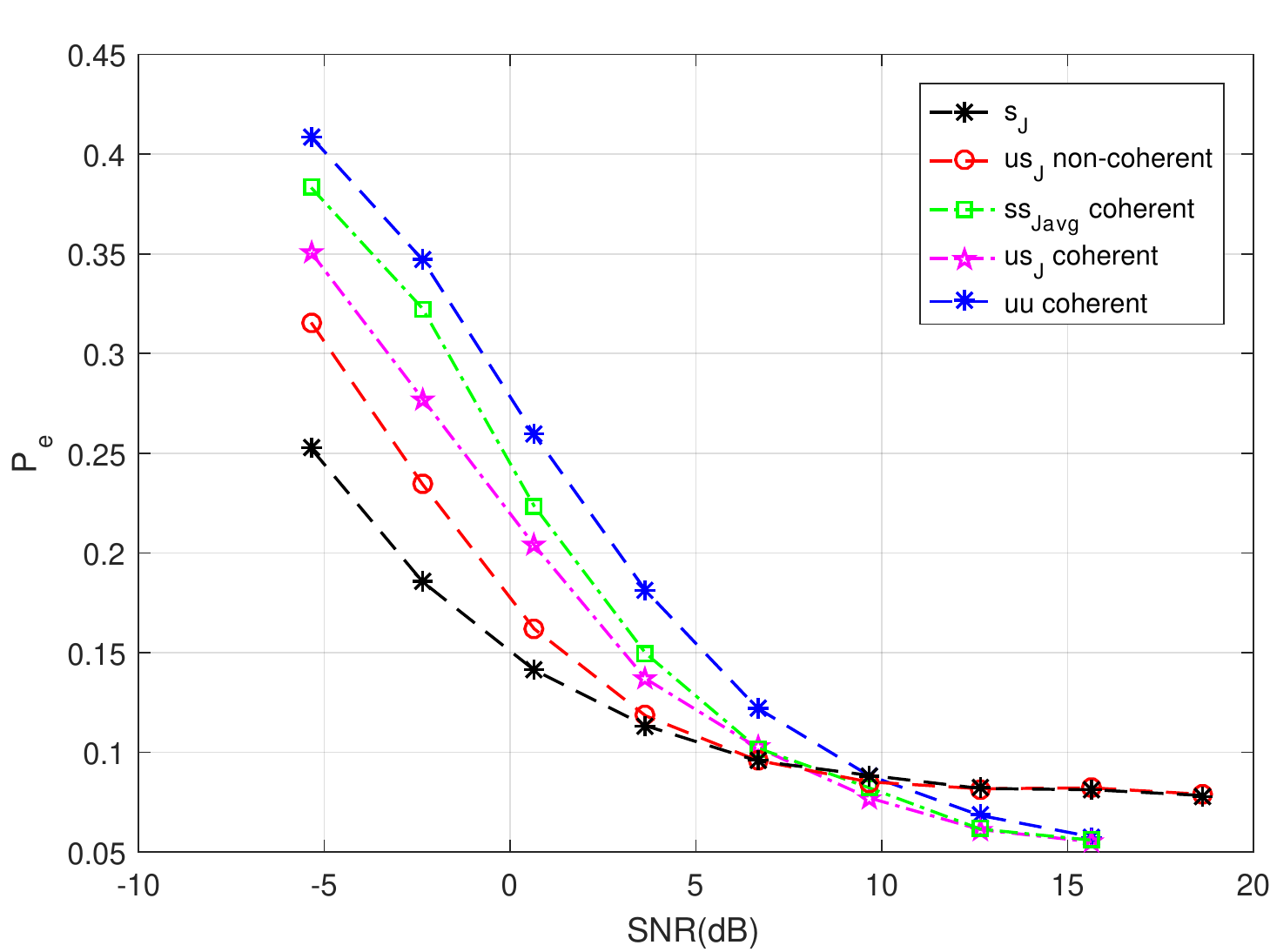}
          \vspace{-0.6cm}
          \centerline{(ii)}\medskip
          \label{fig:x2}
        \vspace{0.4cm}
        \end{minipage}
        \centering
        \begin{minipage}[htp]{0.45\textwidth}
          \centering
          \includegraphics[width=1\textwidth]{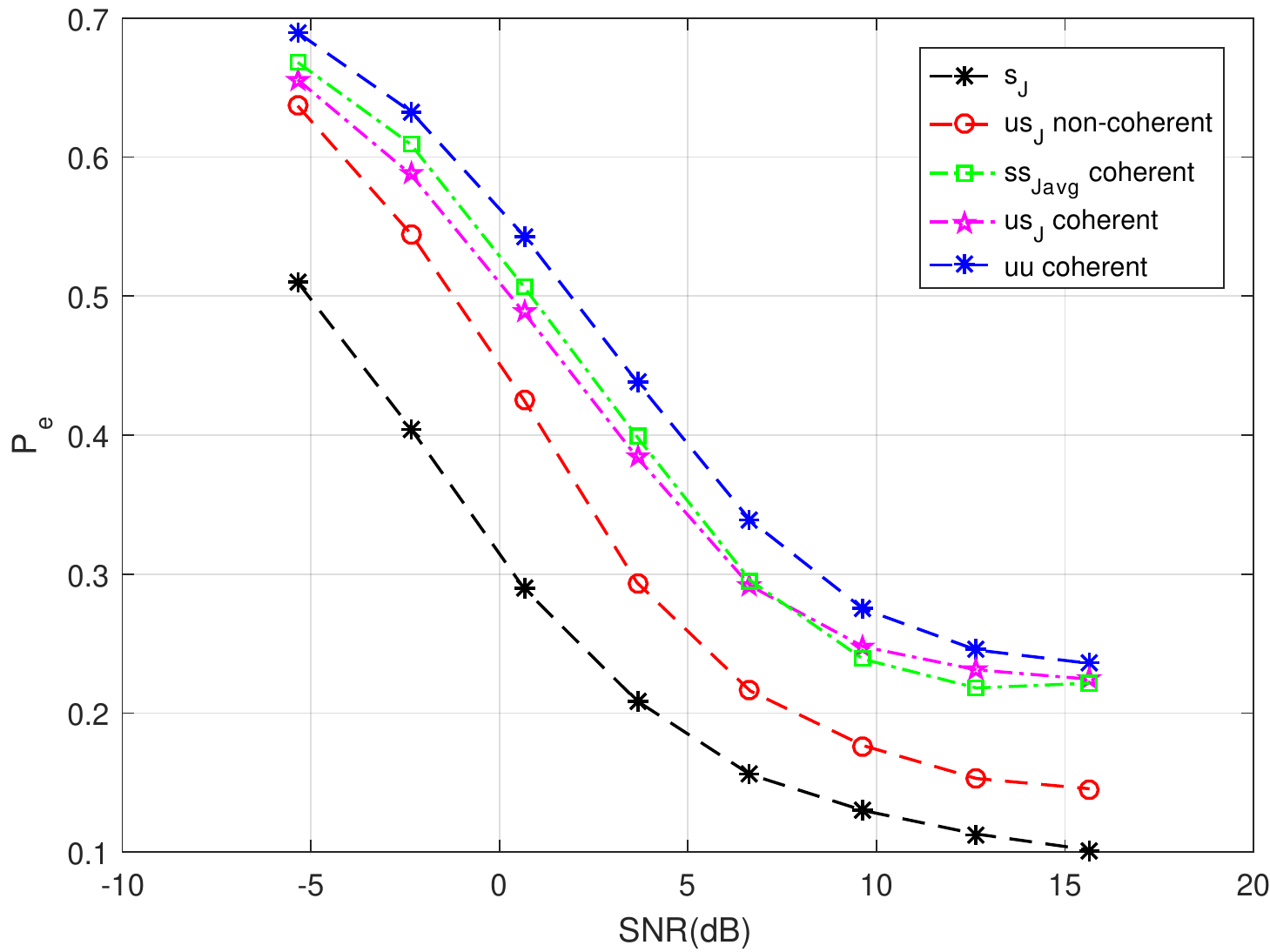}
          
          \vspace{-0.6cm}
          \centerline{(iii)}\medskip
        \vspace{0.4cm}
        \end{minipage}
          \centering
        \begin{minipage}[htp]{0.45\textwidth}
          \centering
          \includegraphics[width=1\textwidth]{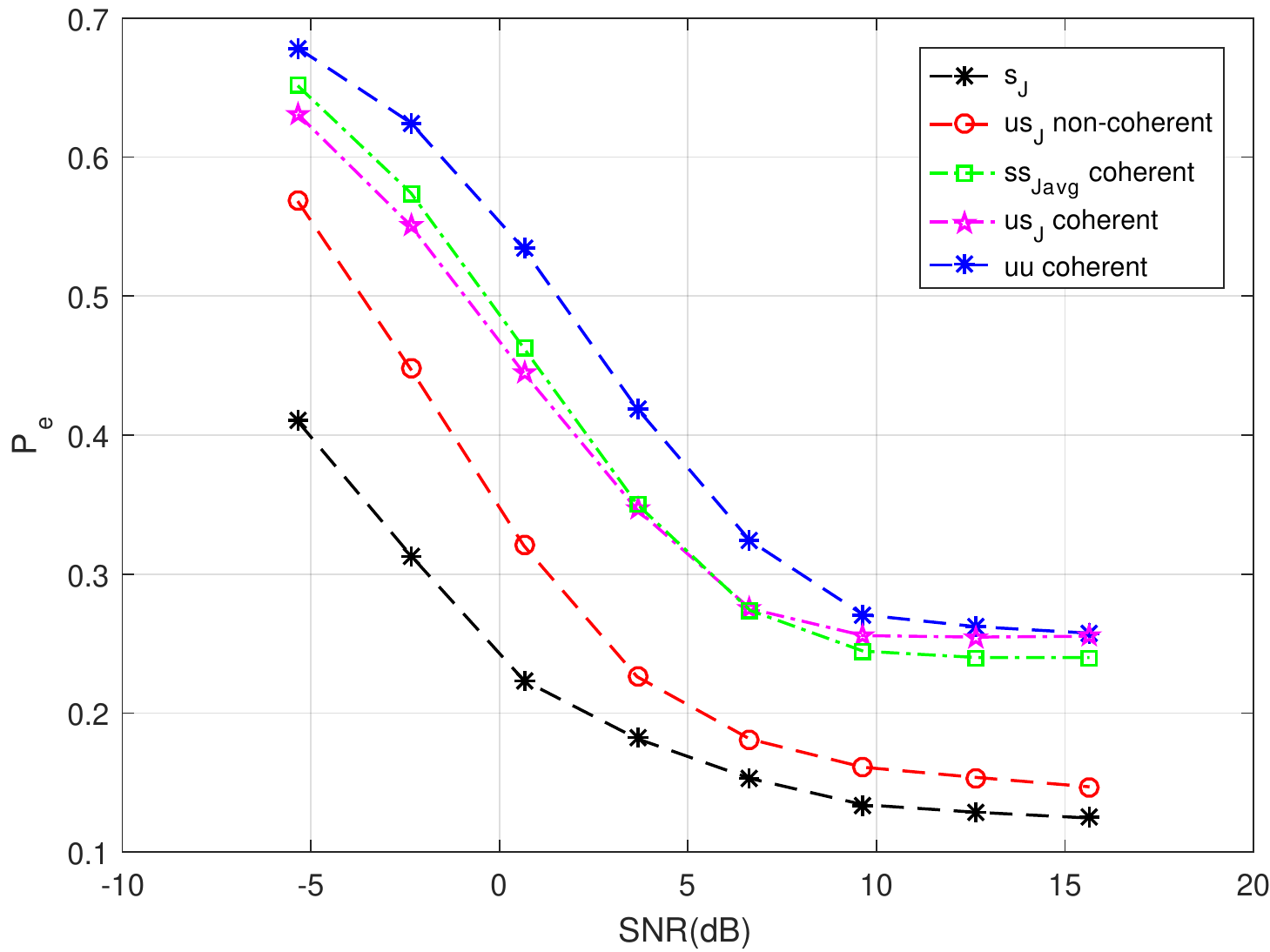}
          \vspace{-0.6cm}
          \centerline{(iv)}\medskip
          \label{fig:x2}
        \vspace{0.4cm}
        \end{minipage}
        
                \caption{Probability of error vs SNR when sensors are situated in equal distance from FC with different performance, (i) M=2, N=5, (ii) M=2, N=10 iii) M=4, N=5, iv) M=4, N=10.}
                \end{figure*}
                
                \begin{figure*}[t!]\label{fig:eqperf}
        \centering
        \begin{minipage}[htp]{0.45\textwidth}
          \centering
          \includegraphics[width=1\textwidth]{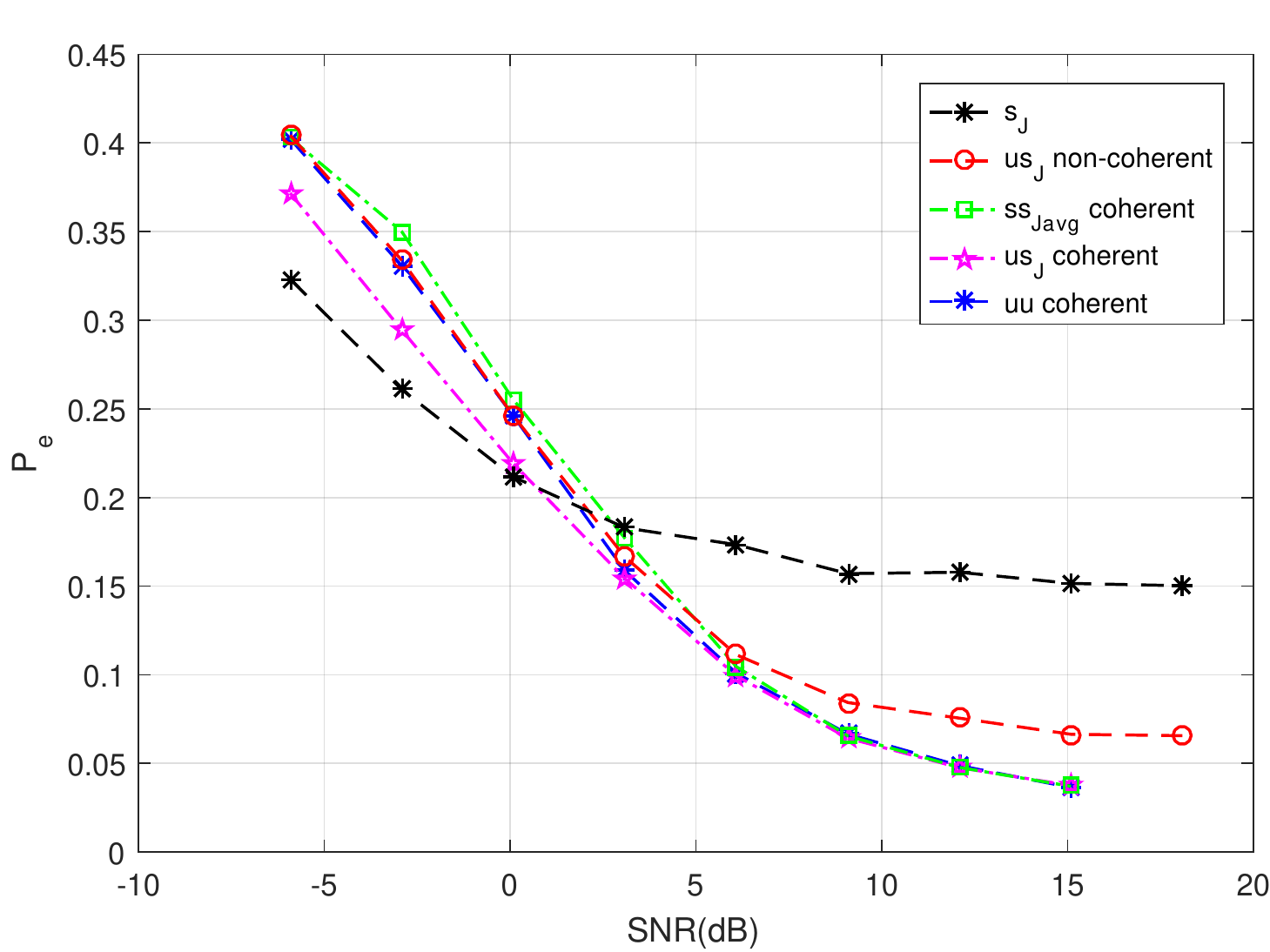}
          \vspace{-0.6cm}
          \centerline{(i)}\medskip
        \vspace{0.4cm}
        \end{minipage}
          \centering
        \begin{minipage}[htp]{0.45\textwidth}
          \centering
          \includegraphics[width=1\textwidth]{eqperf_bin_5.pdf}
          \vspace{-0.6cm}
          \centerline{(ii)}\medskip
          \label{fig:x2}
        \vspace{0.4cm}
        \end{minipage}
        \centering
        \begin{minipage}[htp]{0.45\textwidth}
          \centering
          \includegraphics[width=1\textwidth]{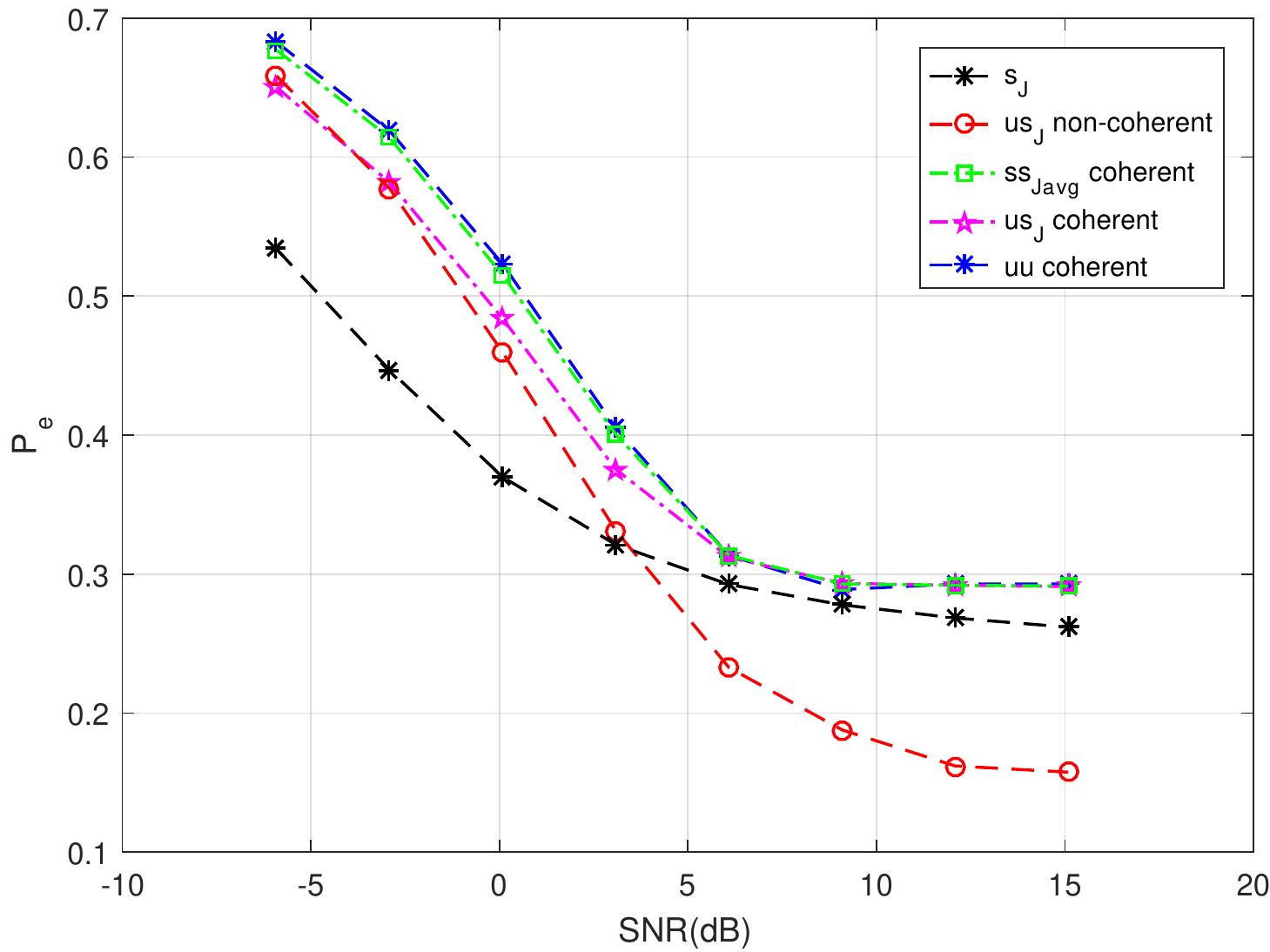}
          \vspace{-0.6cm}
          \centerline{(iii)}\medskip
        \vspace{0.45cm}
        \end{minipage}
          \centering
        \begin{minipage}[htp]{0.45\textwidth}
          \centering
          \includegraphics[width=1\textwidth]{eqperf_4ary_5.pdf}
          \vspace{-0.6cm}
          \centerline{(iv)}\medskip
        \vspace{0.45cm}
        \end{minipage}

                \caption{Probability of error vs SNR  when sensors have equal probability of error but in different distance from  sensors (i) M=2, N=5 (ii) M=2, N=10 iii) M=4, N=5, iv) M=4, N=10.}
                \end{figure*}
                
                \begin{figure*}[t!]\label{fig:clos_rd}
        \centering
        \begin{minipage}[htp]{0.42\textwidth}
          \centering
          \includegraphics[width=1\textwidth]{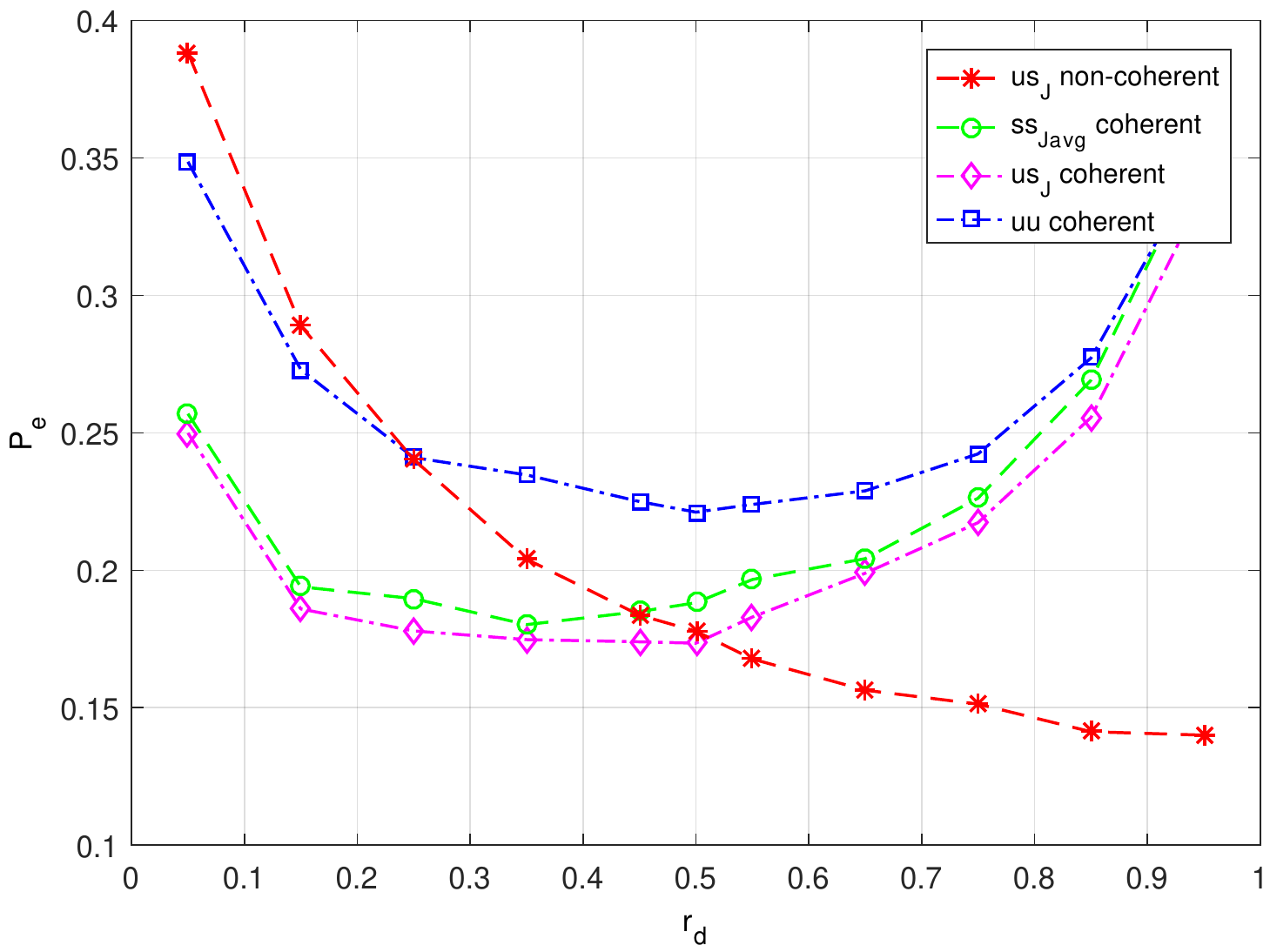}
          \vspace{-0.6cm}
          \centerline{(i)}\medskip
        \vspace{0.4cm}
        \end{minipage}
          \centering
                 \begin{minipage}[htp]{0.42\textwidth}
          \centering
          \includegraphics[width=1\textwidth]{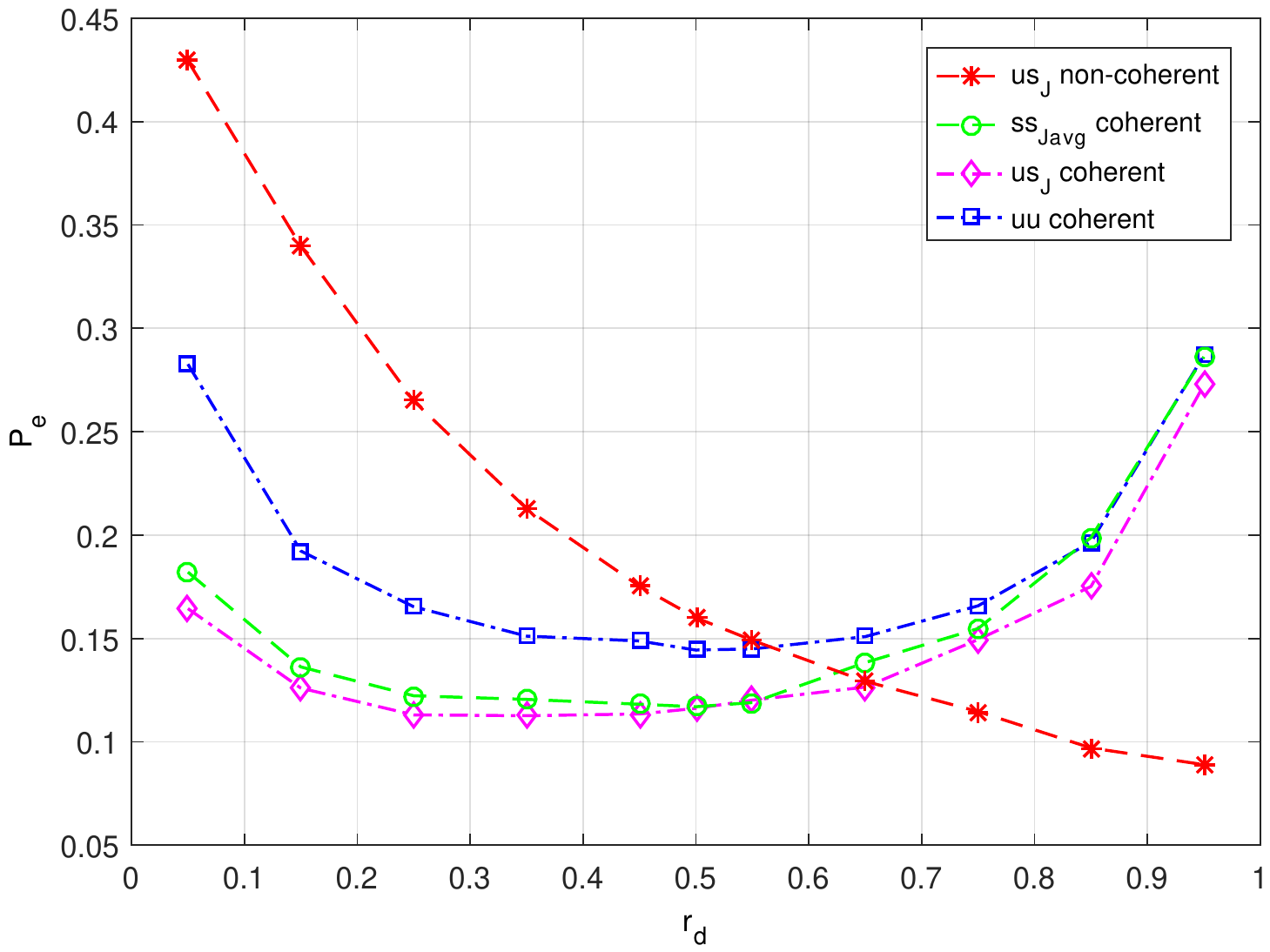}
          \vspace{-0.6cm}
          \centerline{(iii)}\medskip
        \vspace{0.4cm}
        \end{minipage}
                \caption{Probability of error vs $r_d$ for $M=2$ when sensors closer to FC have less probability of error, (i) N=5 (ii) N=10.}
                \end{figure*}
        \begin{figure*}[t!]\label{fig:far_rd}
        \centering
        \begin{minipage}[htp]{0.42\textwidth}
          \centering
          \includegraphics[width=1\textwidth]{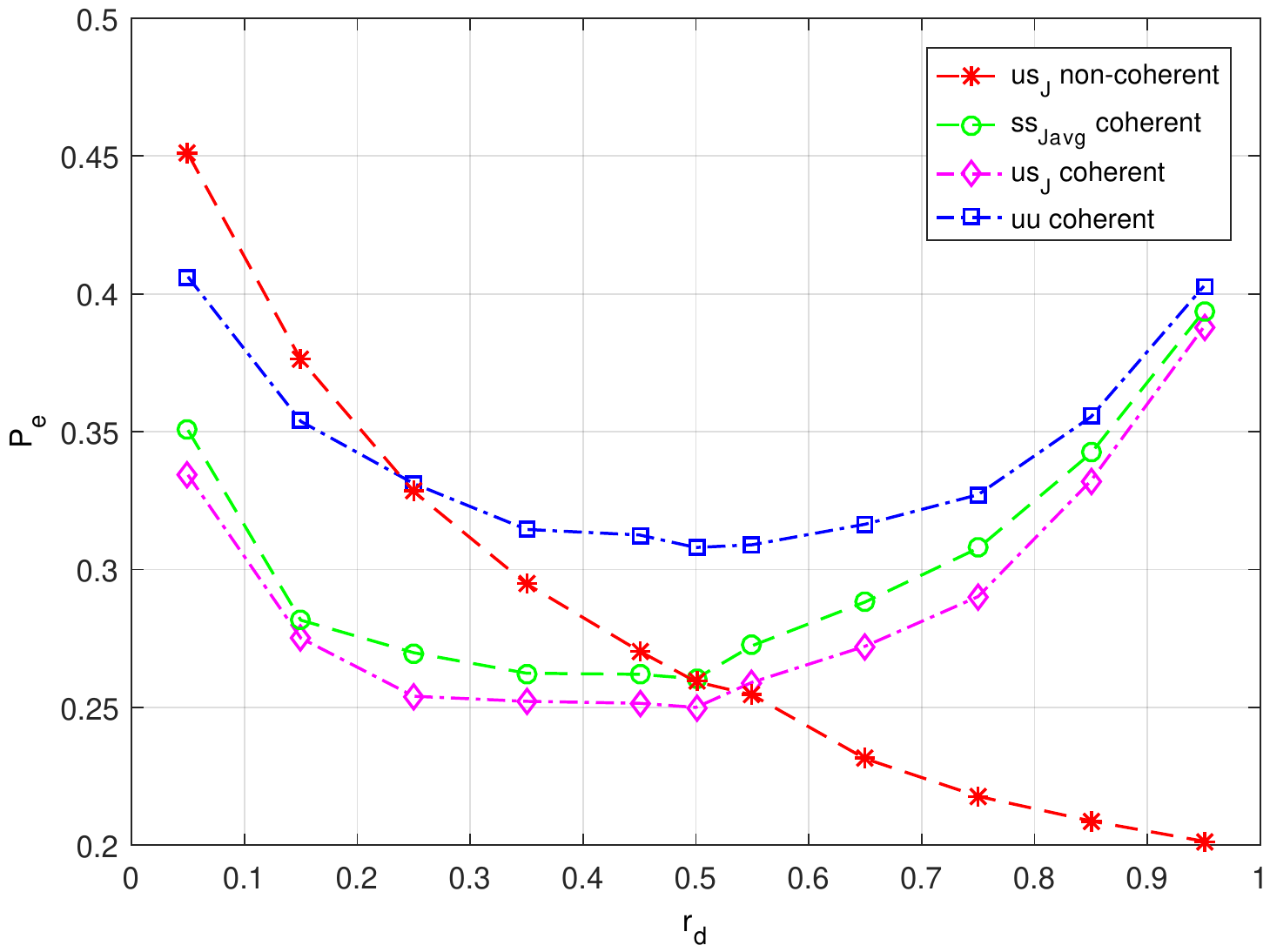}
          \vspace{-0.6cm}
          \centerline{(i)}\medskip
        \vspace{0.4cm}
        \end{minipage}
          \centering
                 \begin{minipage}[htp]{0.42\textwidth}
          \centering
          \includegraphics[width=1\textwidth]{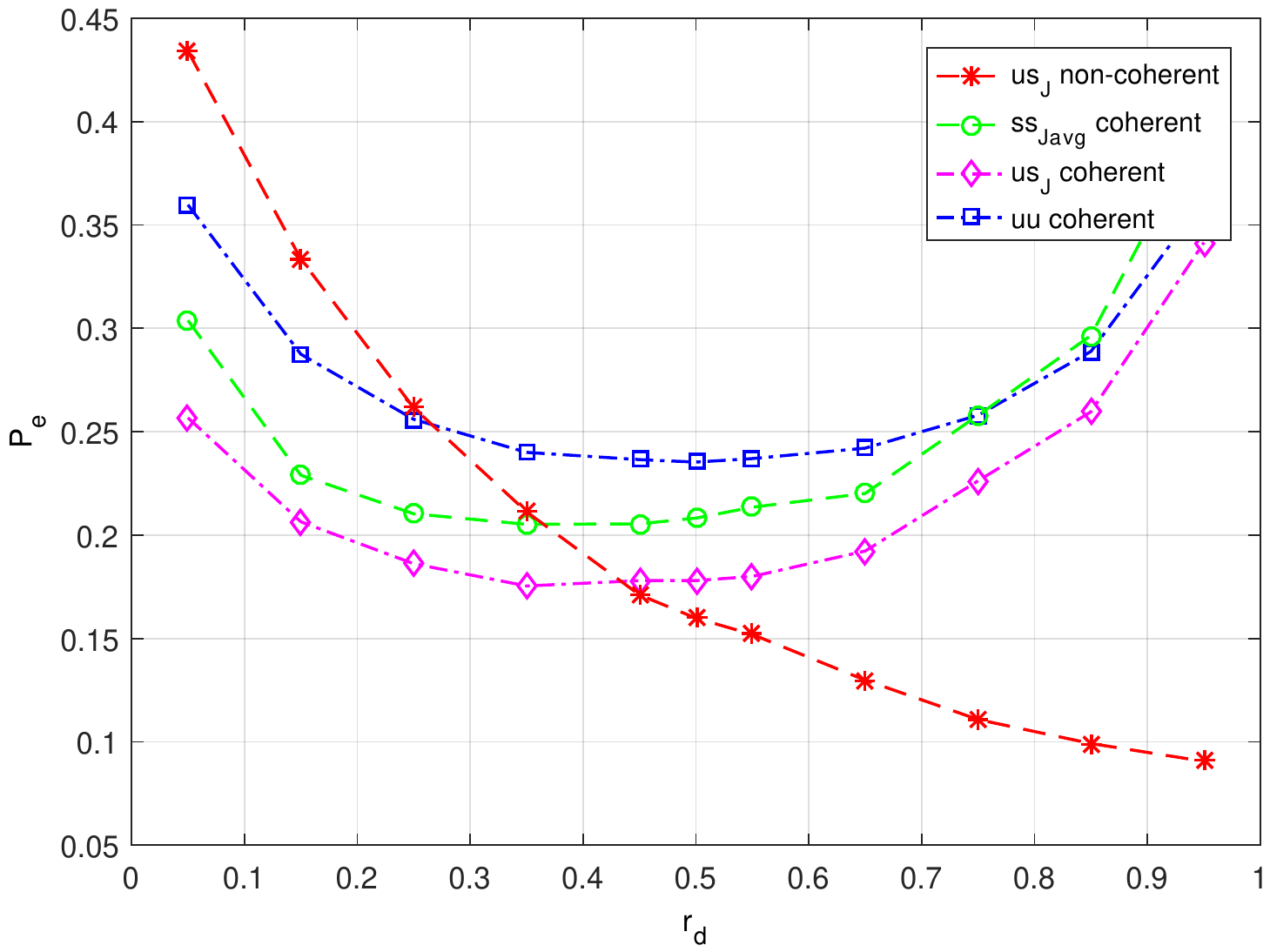}
          \vspace{-0.6cm}
          \centerline{(iii)}\medskip
        \vspace{0.4cm}
        \end{minipage}
                \caption{Probability of error vs $r_d$ for $M=2$ and different receptions when sensors farther from FC have less probability of error, (i) N=5  (ii N=10.}
                \end{figure*}
                \begin{figure*}[t!]\label{fig:eqdist_rd}
        \centering
        \begin{minipage}[htp]{0.42\textwidth}
          \centering
          \includegraphics[width=1\textwidth]{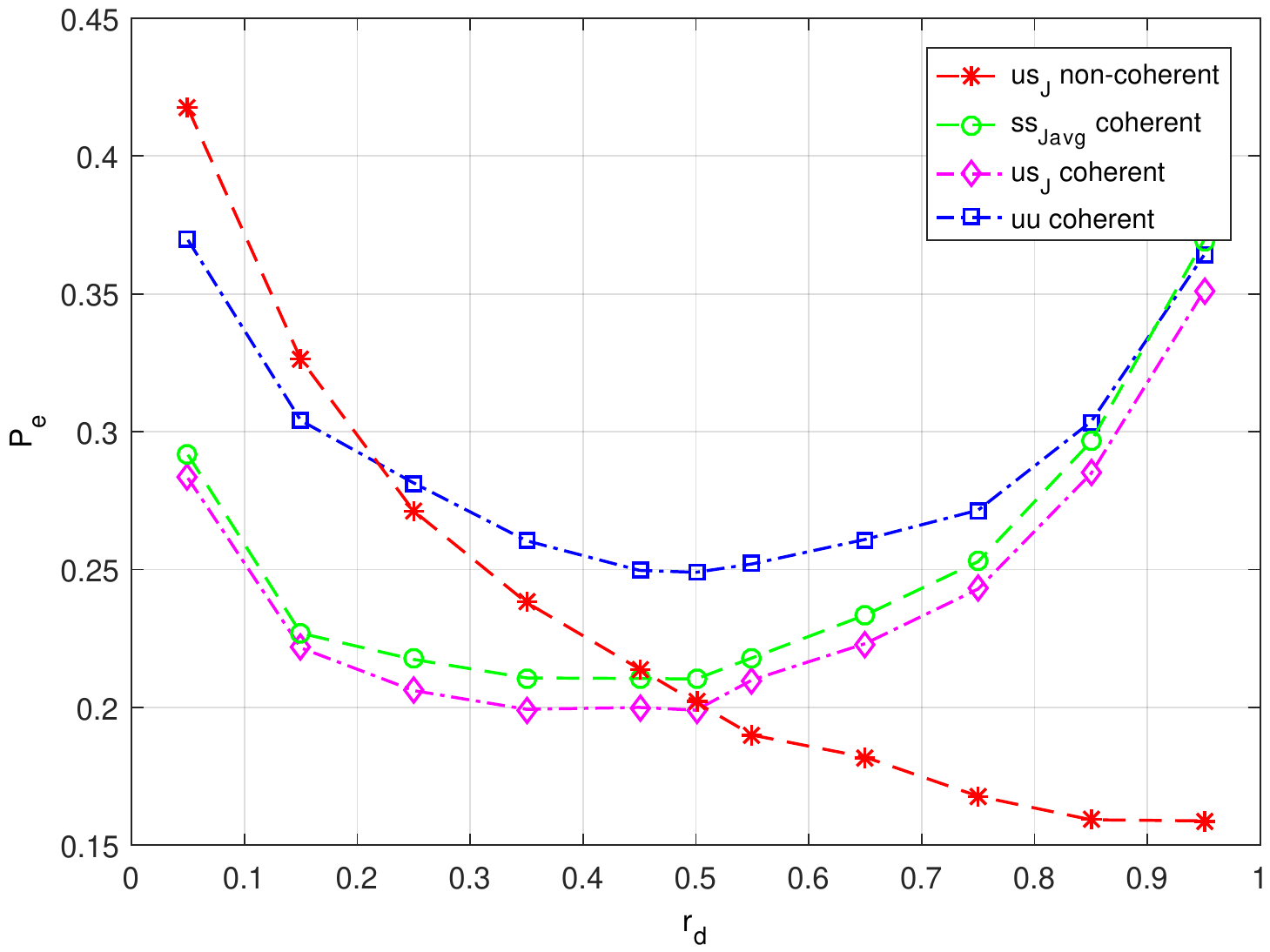}
          \vspace{-0.6cm}
          \centerline{(i)}\medskip
        \vspace{0.4cm}
        \end{minipage}
          \centering
        \begin{minipage}[htp]{0.42\textwidth}
          \centering
          \includegraphics[width=1\textwidth]{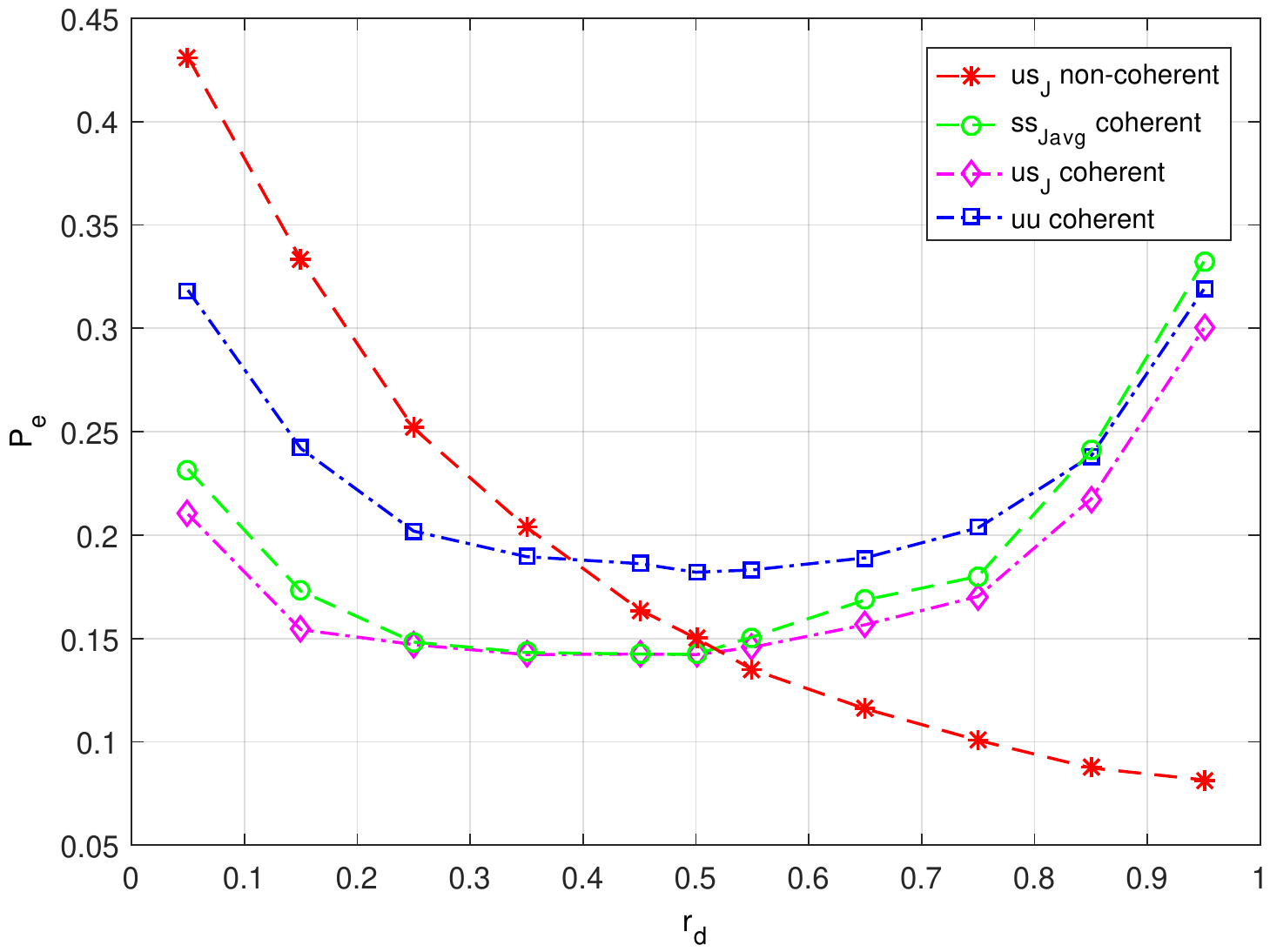}
          \vspace{-0.6cm}
          \centerline{(ii)}\medskip
          \label{fig:x2}
        \vspace{0.4cm}
        \end{minipage}
                \caption{Probability of error vs $r_d$ for $M=2$ and different receptions when sensors are situated in equal distance from FC with different performance, (i) N=5, (ii) N=10.}
                \end{figure*}
                
                \begin{figure*}[t!]\label{fig:eqperf_rd}
        \centering
        \begin{minipage}[htp]{0.42\textwidth}
          \centering
          \includegraphics[width=1\textwidth]{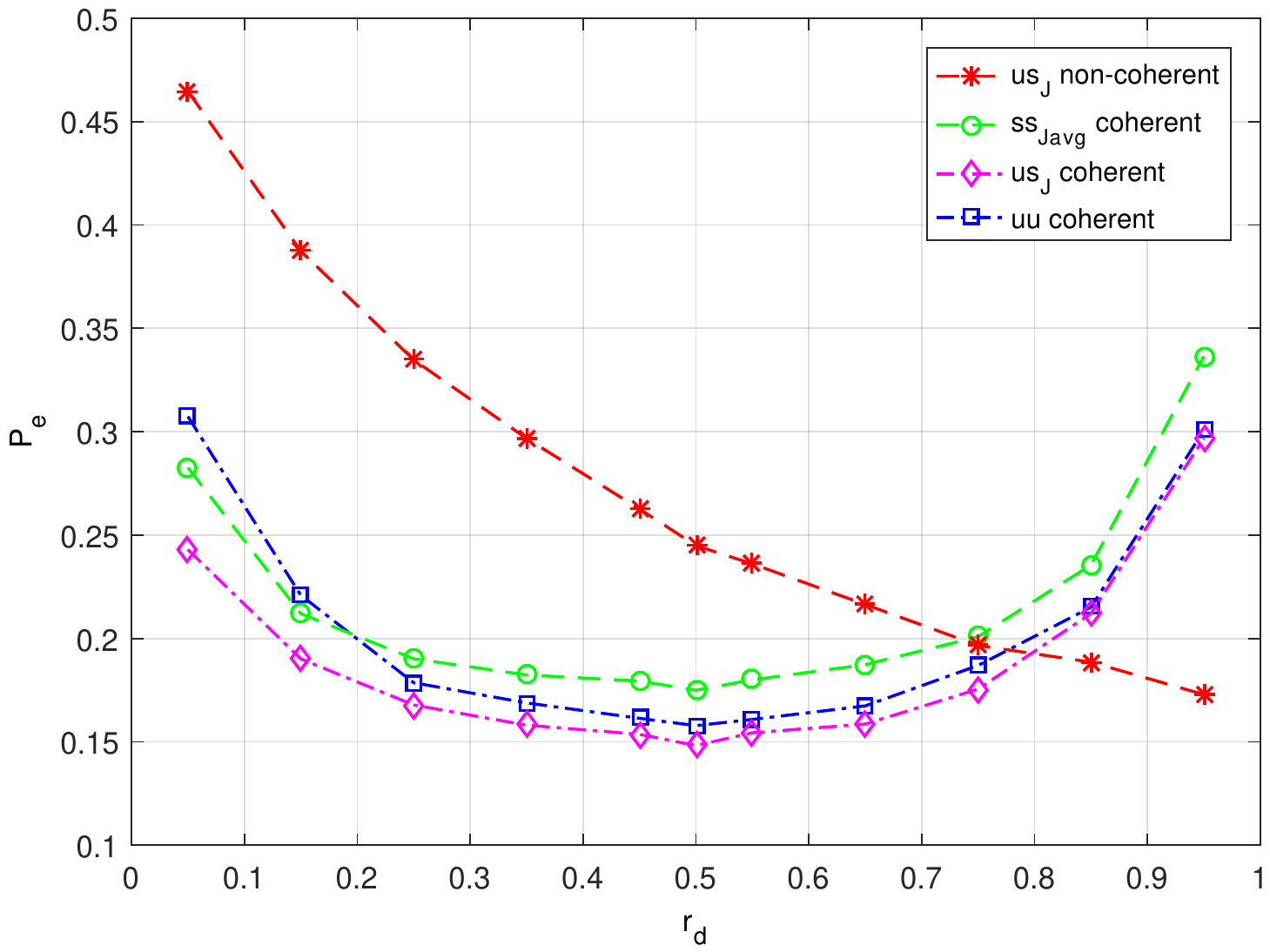}
          \vspace{-0.6cm}
          \centerline{(i)}\medskip
        \vspace{0.4cm}
        \end{minipage}
          \centering
        \begin{minipage}[htp]{0.42\textwidth}
          \centering
          \includegraphics[width=1\textwidth]{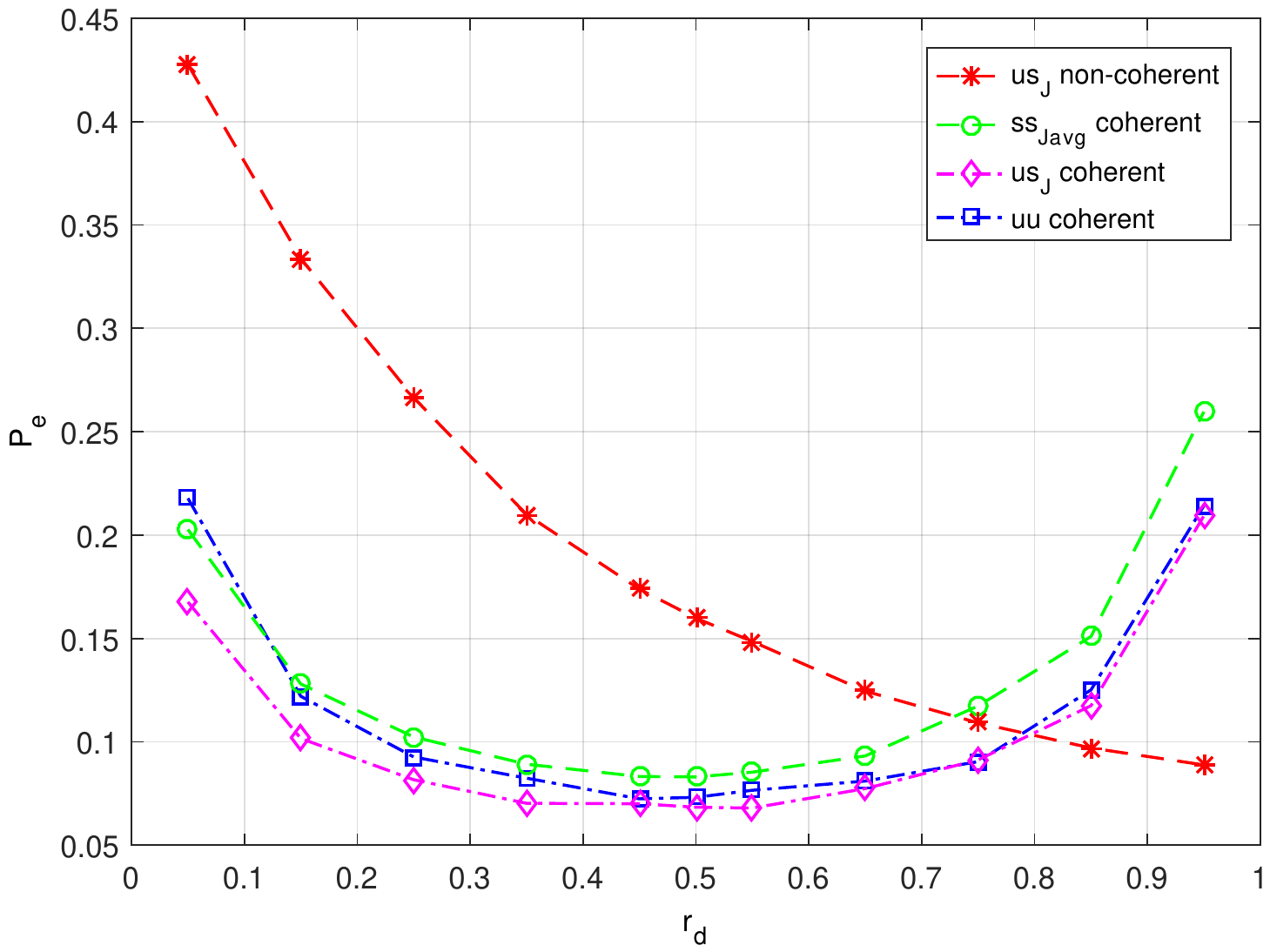}
          \vspace{-0.6cm}
          \centerline{(ii)}\medskip
          \label{fig:x2}
        \vspace{0.4cm}
        \end{minipage}
                \caption{Probability of error vs $r_d$ for $M=2$ when sensors have equal probability of error but in different distance from FC (i) N=5  (ii) N=10.}
                \end{figure*}
        In this section, numerical results are provided to illustrate
        the power allocation scheme developed in this paper. In the
        simulations, we consider the following settings. There are $N$
        sensors scattered around a FC and the distances from the
        sensors to the FC are $\left\lbrace d_k\right\rbrace_{k=1}^{N}$ which we consider to be $d_k=2+2(k-1)$ for $N=5$ and $d_k=k$ for $N=10$. The path loss of signal power
        at the FC from sensor k follows the Motley-Keenan path loss
        model (expressed in dB) without the wall and floor attenuation
        factor \cite{Motley}:
        $PL_k = PL_0 + 10nlog10(d_k/d0)$
        where $PL_0$ is a constant set to $55$dB, and $d_0$ is also a constant
        set to $1m$ in the simulations. Here, $n$ is the path loss exponent,
        which is set to $2$ for free space propagation. The variance of channel from sensor $k$ to FC is $\sigma^2_{h_k} =-PLk$. The noise
        variance from sensors to FC is $\sigma^2_n = −30dBm$ dBm. We define the SNR as $SNR=10log_{10}\frac{P_{tot}\times \sum_k \sigma_{h_k}^2}{N^2\sigma_n^2}$. The FC make the final decision rule based on Bayesian criterion rule which depends on the performance of sensors and their quality of channels. The performance of sensors
        may vary according to their local observation qualities.
        We will consider three receivers: 1) Coherent reception with channel estimation . 2) Non coherent reception with amplitude estimation. 3) Non coherent reception with channel statistic.
        In general, We will investigate 4 different cases:
        \newline Case V-A1:  Sensors closer to FC have less probability of error.
        \newline Case V-A2:  Sensors farther from FC have less probability of error.
       \newline  Case V-A3: All  Sensors have the same distance from FC.
       \newline Case V-A4: All Sensors have the same probability of error but in different distance from FC.
       \newline The probability of error for decision of sensor $k$ after its observation is: $P_e^{s}(k)$: $\sum_{m=1}^M\sum_{\substack{
   i \neq m \\
   1 \leq i \leq M
  }} \pi_m p_{im}^k$.  If $N=5$, we consider  $P_e^{s}=[0.5, 0.5, 0.4, 0.3, 0.1]$. If $N=10, M=2$  $P_e^{s}=[ 0.5 ,0.5, 0.48, 0.46 ,0.4, 0.35, 0.3, 0.2, 0.15, 0.1]$ and For $M=4, N=10$, $P_e^{s}= [0.75, 0.74, 0.7, 0.68, 0.6, 0.55, 0.45, 0.3, 0.2,0.1] $.
        Fig. 1 represents probability of error vs SNR when sensors closer to fusion center have less probability of error. As we can see, in coherent reception, we can reduce total power 2dB to have the same probability of error with power allocation based on total J divergence or based on average J divergence. Also, we can see probability of error for two systems: power allocation based on average J divergence and power allocation based on conditional J divergence are approximately the same which means even though we get feedback from FC to sensors to become aware of channel coefficients for power allocation, the performance will not be changed. So, we can do power allocation without feedback. In non coherent reception with channel estimation, with our power allocation, the amount of power will be reduced around 5dB and 7dB for non coherent reception with amplitude estimation and non coherent reception statistic, respectively. As we can see in the figure, in non coherent statistic receiver, for SNR's greater than 8dB, the probability of error is less for uniform power allocation. Regarding comparing which receiver perform better, The result in this case shows that non coherent statistic perform better for N=5. For N=10, in SNR's less than 8dB non coherent statistic receiver with power allocation based on total J divergence perform better. In higher SNR's, coherent receiver and non coherent statistic receiver with uniform power allocation have the best performance.
        Fig. 2 and Fig. 3 represents probability of error vs SNR for the cases when sensors farther from FC perform better and the case when sensors are situated in equal distance from FC(2m). In these case, we can see similar results to the case when sensors closer to FC perform better. One difference is that in this case the probability of error for coherent reception and power allocation based on average J divergence is worth than power allocation based on conditional J divergence.
        Fig 4. Represents probability of error vs SNR for the case when sensors have equal performance but they are in different distance  from FC. As we can see, our scheme for power allocation, its performance is close to uniform power allocation which indicates the dependency of power allocation to performance of sensors.
        Fig.5 through Fig.8 represent probability of error vs $r_t$ for four different cases. The results show that the optimal $r_t$ for coherent reception is 0.5 and 1 for non coherent reception which means in coherent reception, we should divide power equally between data and training and in non coherent reception, we should devote all power to data.
        

\section{Conclusion}
In summary, we considered a distributed detection wireless
system that is tasked with solving an M-ary hypothesis testing
problem. We studied the effect of wireless channel uncertainty,
due to channel estimation error, on the design and performance
of this system, assuming the sum of transmit powers of
training and data symbols is fixed. In particular, we provided
the optimal MAP fusion rules for training and non-training
based systems. Then, we have developed a power allocation scheme to
distribute the total training and data power budget among the sensors so that the
probability of error at the FC is minimized in terms of the conditional J-divergence and average divergence for different receivers.  Our analytic results show that average divergence for coherent reception will be maximized when the transmit power is equally distributed
between training and data symbols. The simulation results also show that the error probability of this system is minimized when the transmit power is equally distributed
between training and data symbols.However, when the sensors employ PSK modulation along
with coherent reception at the FC the error probability is
minimized when the transmit power is equally distributed
between training and data symbols.
\appendices
\section{}\label{sec:lemma1}
The proof of lemma 1 is as below:
\begin{eqnarray*}
\frac{\partial}{\partial P_{dk}}
\!\!\! J_{tot|\hat{\boldsymbol{G}}}&=
 \sum_{i=1}^{M}\sum_{j=1}^{M}\frac{\hat{\boldsymbol{g}}_k\sigma^2_n\left( \gamma_{ij}^k-B_k\left(j \right) \right) }{\left( \sigma^2_n+\sigma^2_{\tilde{h}_k}P_{dk}+\hat{\boldsymbol{g}}P_{dk}B_i\left(k \right)\right)^2}\\
&+ \frac{\hat{\boldsymbol{g}}_k\sigma^2_n\left( \gamma_{ji}^k-B_k\left(i \right) \right) }{\left( \sigma^2_n+\sigma^2_{\tilde{h}_k}P_{dk}+\hat{\boldsymbol{g}}_kP_{dk}B_j\left(k \right)\right)^2}
\end{eqnarray*}
After some calculation, the term inside the sum will become:
\begin{eqnarray*}
&\!\!\!\frac{\sigma^2_n\hat{\boldsymbol{g}}_k\left( B_k(j)-B_k(i)\right) ^2\left(2\sigma^2_n+2\sigma^2_{\tilde{h}_k}P_{dk}+\hat{\boldsymbol{g}}_kP_{dk}\left( B_j(j)+B_k(i)\right)  \right) }{\left( \sigma^2_n+\sigma^2_{\tilde{h}_k}P_{dk}+\hat{\boldsymbol{g}}_kP_{dk}B_j\left(k \right)\right)^2\left( \sigma^2_n+\sigma^2_{\tilde{h}_k}P_{dk}+\hat{\boldsymbol{g}}_kP_{dk}B_i\left(k \right)\right)^2}\\
&\!\!\!+\frac{|B_i(k)-B_j(k)|^2}{\left( \sigma^2_n+\sigma^2_{\tilde{h}_k}P_{dk}+\hat{\boldsymbol{g}}_kP_{dk}B_i\left(k \right)\right)^2} +\frac{|B_i(k)-B_j(k)|^2}{\left( \sigma^2_n+\sigma^2_{\tilde{h}_k}P_{dk}+\hat{\boldsymbol{g}}_kP_{dk}B_j\left(k \right)\right)^2}\geq 0
\end{eqnarray*}
Therefore $\frac{\partial}{\partial P_{dj}}
J_{tot|\hat{\boldsymbol{G}}}(P_{d1}, · · · , P_{dN})|_{P_{dk}\geq 0}\geq 0$.
\newline The first order derivative of objective function respect to $P_{tj}$ is $\frac{\partial J_{tot|\hat{\boldsymbol{G}}}}{\partial P_{tj}}=\frac{\partial J_{tot|\hat{\boldsymbol{G}}}}{\partial \boldsymbol{\hat{g}}_k}\frac{\partial \boldsymbol{\hat{g}}_k}{\partial P_{tj}}+\frac{\partial J_{tot|\hat{\boldsymbol{G}}}}{\partial \sigma_{w_k}^2}\frac{\partial \sigma_{w_k}^2}{\partial P_{tj}}$. Similarly, we can prove that $\frac{\partial J_{tot|\hat{\boldsymbol{G}}}}{\partial \boldsymbol{\hat{g}}_k} \geq 0$ and $\frac{\partial J_{tot|\hat{\boldsymbol{G}}}}{\partial \sigma_{w_k}^2} \leq 0$. Since $\frac{\partial \boldsymbol{\hat{g}}_k}{\partial P_{tj}} \geq 0$ and $\frac{\partial \sigma_{w_k}^2}{\partial P_{tj}} \leq 0$, we can conclude $\frac{\partial J_{tot|\hat{\boldsymbol{G}}}}{\partial P_{tj}} \geq 0$.
\section{}\label{sec:lemma2}
The proof of lemma 2 is as below:
\newline The first order derivative of $J_{avg}$ respect to $r_k$ is:
\begin{eqnarray*}
\frac{\partial J_{avg}}{\partial r_k}&=& \sum_{i=1}^{M}\sum_{j=1}^{M}\left(1-\frac{\gamma_{ji}^k}{B_i(k)}\right) \frac{\partial x_k}{\partial r_k}\frac{\partial D\left(\frac{x_k}{B_i(k)}\right)}{\partial x_k} \\
&&+\left(1-\frac{\gamma_{ij}^k}{B_j(k)}\right)\frac{\partial x_k}{\partial r_k}\frac{\partial D\left(\frac{x_k}{B_i(k)}\right)}{\partial x_k} 
\end{eqnarray*}
Now, We prove that $D(x)$ is an increasing function of $x$. We show that $D(x_2)\geq D(x_1)$ if $x_2\geq x_1$. 
\begin{eqnarray*}
D(x_2)-D(x_1)&=& x_2e^{x_2}\int_{x_2}^{\infty}\frac{e^{-t}}{t}dt-x_1e^{x_1}\int_{x_1}^{\infty}\frac{e^{-t}}{t}dt\\
&&\int_{0}^{\infty}e^{-t}\left( \frac{x_2}{t+x_2}-\frac{x_1}{t+x_2}\right)dt\\
&&\int_{0}^{\infty}e^{-t}\left( \frac{tx_2-tx_1}{\left( t+x_2\right) \left( t+x_1\right) }\right)dt \geq 0\\
\end{eqnarray*}
Therefore, $\frac{\partial D(x)}{\partial x} \geq 0$.
\newline  Based on the assumption of  $\left(1-\frac{\gamma_{ji}^k}{B_i(k)}\right)<0$ and $\frac{\partial x_k}{\partial r_k}$, we will have:
\begin{equation*}
\begin{cases}
     \frac{\partial J_{avg}}{\partial r_k}>0 , & \text{if}\ r_k<\frac{1}{2} \\
     \frac{\partial J_{avg}}{\partial r_k}=0 , & \text{if}\ r_k=\frac{1}{2} \\
      \frac{\partial J_{avg}}{\partial r_k}<0, & \text{if}\ r_k>\frac{1}{2}
    \end{cases}
\end{equation*}
Based on above relationship, when $r_k<\frac{1}{2}$, $J_{avg}$ is an increasing function of $r_k$ and for $r_k > \frac{1}{2}$, $J_{avg}$ is a decreasing function of $r_k$. Therefore, we can conclude the optimal $r_k^*=0.5$.
\section{}\label{sec:lemma3}
The proof of lemma 3 is as below:
\newline The first derivative of $J_{avg}$ respect to $P_k$ is
\begin{equation}\label{partial derivative}
\sum_{i=1}^{M}\sum_{j=1}^{M}\frac{\partial f_k}{\partial x_k}\frac{\partial x_k}{\partial P_k}=\sum_{i=1}^{M}\sum_{j=1}^{M}\left( \frac{\partial f_k^1}{\partial x_k}+\frac{\partial f_k^2}{\partial x_k}\right) \frac{\partial x_k}{\partial P_k}
\end{equation}
 Where $f_k$ is:
\begin{eqnarray*}
\label{f_k}
&& \!\!\!\!\!\!\!\!\!\!\!f_k=\sum_{i=1}^{M}\sum_{j=1}^{M}\left[ \frac{\gamma_{ji}^k}{B_i(k)}+\left(1-\frac{\gamma_{ji}^k}{B_i(k)} \right) D(\frac{x_k}{B_i(k)})\right] \\
&&\left[ \sum_{i=1}^{M}\sum_{j=1}^{M}+\frac{\gamma_{ij}^k}{B_j(k)}+\left(1-\frac{\gamma_{ij}^k}{B_j(k)} \right)D(\frac{x_k}{B_j(k)})\right]  
\end{eqnarray*}
 The derivative of first bracket respect to $P_k$ become:
\begin{equation}\label{first derivative f1}
\sum_{i=1}^{M}\sum_{j=1}^{M}\left( 1-\frac{\gamma_{ji}^k}{B_i(k)}\right) \frac{\partial D}{\partial x_k}\frac{\partial x_k}{\partial P_k}
\end{equation}
Since $\partial D(x)/\partial(x)\geq 0$, $1-\frac{\gamma_{ji}^k}{B_i(k)} < 0$, $\partial x_k/\partial P_k \leq 0$, we conclude that The derivative of first bracket is  non negative. Similarly, We can see that derivative of second bracket in (\ref{f_k}) is non negative. Therefore, We conclude that $\partial J_{avg}/\partial P_k \geq 0$. 


\begin{thebibliography}{10}

\bibitem{biaochen2006}
Ruixin Niu, Biao Chen, and P.K. Varshney.
\newblock Fusion of decisions transmitted over rayleigh fading channels in
  wireless sensor networks.
\newblock {\em Signal Processing, IEEE Transactions on}, 54(3):1018--1027,
  March 2006.
  
 \bibitem{Jiang}
F.~Jiang, J.~Chen, A.~L. Swindlehurst, and J.~A. López-Salcedo.
\newblock Massive mimo for wireless sensing with a coherent multiple access
  channel.
\newblock {\em IEEE Transactions on Signal Processing}, 63(12):3005--3017, June
  2015.

\bibitem{Varsh}
H.~He and P.~K. Varshney.
\newblock Fusing censored dependent data for distributed detection.
\newblock {\em IEEE Transactions on Signal Processing}, 63(16):4385--4395, Aug
  2015.

\bibitem{Braca}
P.~Braca, S.~Marano, and V.~Matta.
\newblock Single-transmission distributed detection via order statistics.
\newblock {\em IEEE Transactions on Signal Processing}, 60(4):2042--2048, April
  2012.

\bibitem{Nad}
V.~S.~S. Nadendla and P.~K. Varshney.
\newblock Design of binary quantizers for distributed detection under secrecy
  constraints.
\newblock {\em IEEE Transactions on Signal Processing}, 64(10):2636--2648, May
  2016.
  
\bibitem{vincentpoor}
Xin Zhang, H.V. Poor, and Mung Chiang.
\newblock Optimal power allocation for distributed detection over mimo channels
  in wireless sensor networks.
\newblock {\em Signal Processing, IEEE Transactions on}, 56(9):4124--4140, Sept
  2008.
  
  
\bibitem{multipleantennamac}
M.K. Banavar, A.D. Smith, C.~Tepedelenlioglu, and A.~Spanias.
\newblock On the effectiveness of multiple antennas in distributed detection
  over fading macs.
\newblock {\em Wireless Communications, IEEE Transactions on},
  11(5):1744--1752, May 2012.

\bibitem{channelestimation}
A.~Vosoughi and Yupeng Jia.
\newblock How does channel estimation error affect average sum-rate in two-way
  amplify-and-forward relay networks?
\newblock {\em Wireless Communications, IEEE Transactions on},
  11(5):1676--1687, May 2012.

\bibitem{6516873}
H.R. Ahmadi and A.~Vosoughi.
\newblock Impact of wireless channel uncertainty upon distributed detection
  systems.
\newblock {\em Wireless Communications, IEEE Transactions on},
  12(6):2566--2577, June 2013.

  
\bibitem{confhamid}
H.R. Ahmadi and A.~Vosoughi.
\newblock Impact of channel estimation error on decentralized detection in
  bandwidth constrained wireless sensor networks.
\newblock In {\em Military Communications Conference, 2008. MILCOM 2008. IEEE},
  pages 1--7, Nov 2008.
  
\bibitem{nevat}
I.~Nevat, G.W. Peters, and I.B. Collings.
\newblock Distributed detection in sensor networks over fading channels with
  multiple antennas at the fusion center.
\newblock {\em Signal Processing, IEEE Transactions on}, 62(3):671--683, Feb
  2014.

\bibitem{pacvsmac}
C.~R. Berger, M.~Guerriero, S.~Zhou, and P.~Willett.
\newblock Pac vs. mac for decentralized detection using noncoherent modulation.
\newblock {\em IEEE Transactions on Signal Processing}, 57(9):3562--3575, Sept
  2009.
  
 \bibitem{Li}
F.~Li, J.~S. Evans, and S.~Dey.
\newblock Decision fusion over noncoherent fading multiaccess channels.
\newblock {\em IEEE Transactions on Signal Processing}, 59(9):4367--4380, Sept
  2011.
  
\bibitem{Ciu}
D.~Ciuonzo, G.~Romano, and P.~Salvo Rossi.
\newblock Optimality of received energy in decision fusion over rayleigh fading
  diversity mac with non-identical sensors.
\newblock {\em IEEE Transactions on Signal Processing}, 61(1):22--27, Jan 2013.

\bibitem{Maleki}
N.~Maleki, A.~Vosoughi, and N.~Rahnavard.
\newblock Distributed binary detection over fading channels: Cooperative and
  parallel architectures.
\newblock {\em IEEE Transactions on Vehicular Technology}, 65(9):7090--7109,
  Sept 2016.

\bibitem{1413463}
Jayesh~H. Kotecha, V.~Ramachandran, and A.M. Sayeed.
\newblock Distributed multitarget classification in wireless sensor networks.
\newblock {\em Selected Areas in Communications, IEEE Journal on},
  23(4):703--713, April 2005.

\bibitem{nahal}
N.~Maleki and A.~Vosoughi.
\newblock Channel-aware m-ary distributed detection: Optimal and suboptimal
  fusion rules.
\newblock In {\em Statistical Signal Processing Workshop (SSP), 2012 IEEE},
  pages 644--647, Aug 2012.
  
  
\bibitem{blindalgorithm}
A.~Jeremic, Kon~Max Wong, and Bin Liu.
\newblock Optimal distributed detection of multiple hypotheses using blind
  algorithm.
\newblock In {\em Acoustics, Speech and Signal Processing, 2009. ICASSP 2009.
  IEEE International Conference on}, pages 2241--2244, April 2009.
  
\bibitem{Nur}
E.~Nurellari, D.~McLernon, and M.~Ghogho.
\newblock Distributed two-step quantized fusion rules via consensus algorithm
  for distributed detection in wireless sensor networks.
\newblock {\em IEEE Transactions on Signal and Information Processing over
  Networks}, 2(3):321--335, Sept 2016.
  
 \bibitem{Maya}
J.~A. Maya, L.~Rey Vega, and C.~G. Galarza.
\newblock Optimal resource allocation for detection of a gaussian process using
  a mac in wsns.
\newblock {\em IEEE Transactions on Signal Processing}, 63(8):2057--2069, April
  2015.


\bibitem{Naj}
M.~Najimi, A.~Ebrahimzadeh, S.~M.~H. Andargoli, and A.~Fallahi.
\newblock Energy-efficient sensor selection for cooperative spectrum sensing in
  the lack or partial information.
\newblock {\em IEEE Sensors Journal}, 15(7):3807--3818, July 2015.

\bibitem{deflection}
Z.~Xu, J.~Huang, and Q.~Zhang.
\newblock Power constrained partially coherent distributed detection over
  fading multiaccess channels.
\newblock {\em IEEE Sensors Journal}, 13(7):2729--2736, July 2013.

\bibitem{goodman}
Hyoung-Soo Kim and N.A. Goodman.
\newblock Power control strategy for distributed multiple-hypothesis detection.
\newblock {\em Signal Processing, IEEE Transactions on}, 58(7):3751--3764, July
  2010.

\bibitem{Mokhtar}
Mokhtar~S. Bazaraa.
\newblock {\em Nonlinear Programming: Theory and Algorithms}.
\newblock Wiley Publishing, 3rd edition, 2013.


\bibitem{cihan}
H.~Senol and C.~Tepedelenlioglu.
\newblock Performance of distributed estimation over unknown parallel fading
  channels.
\newblock {\em IEEE Transactions on Signal Processing}, 56(12):6057--6068, Dec
  2008.
  
\bibitem{Chaudhary}
Muhammad~Hafeez Chaudhary and Luc Vandendorpe.
\newblock Adaptive power allocation in wireless sensor networks with spatially
  correlated data and analog modulation: Perfect and imperfect csi.
\newblock {\em EURASIP J. Wirel. Commun. Netw.}, 2010:12:1--12:14, January
  2010.
  
\bibitem{Motley}
A. J. Motley and J. M. P. Keenan.
\newblock Personal communication radio coverage in buildings at 900 MHz and 1700 MHz.
\newblock {\em Electronics Letters.}, vol. 24, no. 12, pp. 763-764, 9 June 1988.
  

\end{thebibliography}
       \end{document}